\documentclass[aps,twocolumn,showpacs]{revtex4}
\usepackage{amsmath}
\usepackage{bm}
\usepackage{dcolumn}
\usepackage{graphicx}
\begin{document}
\author{Yu.L.~Raikher}
\thanks{Corresponding author. Fax: +7 (3422) 336957; e-mail: raikher@icmm.ru}
\author{V.I.~Stepanov}
\affiliation{Institute of Continuous Media Mechanics, Ural Division of RAS, 614013, %
             Perm, Russia}
\title{Linear and nonlinear superparamagnetic relaxation \\ at high anisotropy barriers}

\begin{abstract}
The micromagnetic Fokker-Planck equation is solved for a uniaxial particle in the
low-temperature limit. Asymptotic series in the parameter that is the inverse barrier
height--to--temperature ratio are derived. With the aid of these series, the expressions for
the superparamagnetic relaxation time and the odd-order dynamic susceptibilities are
presented. The obtained formulas are both quite compact and practically exact in the low (with
respect to FMR) frequency range that is proved by comparison with the numerically-exact
solution of the micromagnetic equation. The susceptibilities formulas contain angular
dependencies that allow to consider textured as well as randomly oriented particle assemblies.
Our results advance the previous two-level model for nonlinear superparamagnetic relaxation.
\end{abstract}

\maketitle

\section{Introduction \label{sec:1}}
The problem of superparamagnetic relaxation in single-domain ferroparticles formulated,
explained and basically analyzed by N\'eel\cite{Neel-AGCR:49} about fifty years ago, since
then keeps to attract attention. Nowadays this interest is mainly due to the expanding number
of nanometer granular magnetic media used in information storage and related high
technologies.

When analyzing magnetic dispersions, solid or fluid, a promising idea is to evaluate the
granulometric content, particle material parameters and relaxation rates by combining the data
on linear and nonlinear dynamic susceptibilities. Since recently, this approach (it originates
from the spin glass science) became quite feasible in experimental
realization.\cite{BiOh-JPSJ:96} However, to benefit from it, one needs an adequate model.
Surprisingly, until nowadays the N\'eel\cite{Neel-AGCR:49} concept of superparamagnetic
behavior of fine magnetic particles that had been substantially advanced by
Brown\cite{Brown-JAP:59,Brown-PR:63} and refined by numerous researchers (see the review
article\cite{DoFi-ACP:97} with about 400 references), lacks a nonlinear extension.

In Ref.~\onlinecite{RaSt-PRB:97} we begun to fill up this gap and proposed a numerical
procedure involving continuous fractions by means of which the linear and cubic
susceptibilities for a solid system of uniaxial fine particles could be obtained. With
allowance for the polydispersity of real samples, the worked out description provided a fairly
good agreement with the dynamic magnetic measurements taken on Co-Cu
nanocomposites.\cite{BiOh-JPSJ:96} Recently, our approach was used
successfully\cite{SpFi-JMMM:01} for the linear and cubic susceptibilities of the samples of
randomly oriented $\gamma$-Fe$_2$O$_3$ nanoparticles. Hereby we carry on the build-up of the
nonlinear superparamagnetic relaxation theory by working out a set of compact and accurate
analytical expressions that considerably facilitate calculations as well as experiment
interpretation.

The paper is arranged in the following way. In Sec.~\ref{sec:2} we discuss the problem of
superparamagnetic relaxation and show the way to obtain the asymptotic solution for the
micromagnetic Fokker-Planck equation in the uniaxial case. In Sec.~\ref{sec:3} the
perturbative expansions for the orientational distribution function are obtained, which are
used in Sec.~\ref{sec:4} to construct asymptotic expressions for the nonlinear dynamic
susceptibilities. The explicit forms of those expansions are given and their accuracy is
proved by comparison with the results of numerical calculations. Sec.~\ref{sec:5} contains the
enveloping discussion.

\section{Superparamagnetic relaxation times \label{sec:2}}
\subsection{Uniaxially anisotropic particle \label{sec:2.1}}
The cornerstone of the superparamagnetic relaxation theory is the Arrhenius-like law for the
relaxation rate of a magnetic moment of a single domain particle predicted by N\'eel in 1949.
The framework of this classical problem is as follows. Consider an immobile (e.g.\ fixed
inside a solid matrix) single-domain grain of a volume $v$. This particle possesses a uniaxial
volume magnetic anisotropy, $K$ being its energy density and $\bm{n}$ its easy axis direction.
Since the temperature $T$ is assumed to be much lower than the Curie point, the particle
magnetization $I$, as a specific parameter, is practically constant and the magnitude of the
particle magnetic moment may be written as $\mu=Iv$. Denoting its direction by a unit vector
$\bm{e}$, one concludes that the magnetic state of such a particle is exhaustively
characterized by a pair of vectors: $\bm{\mu}=Iv\bm{e}$ and $\bm{n}$. Thence, the
orientation-dependent part of the particle energy (in the absence of external magnetic fields)
is
\begin{equation}                                                        \label{eq:01}
U=-Kv(\bm{e}\cdot\bm{n})^2,
\end{equation}
where $K$ is assumed to be positive. As Eq.~(\ref{eq:01}) shows, this energy has two equal
minima. They are separated by the potential barrier of the height $Kv$ and correspond to
$\bm{e}\parallel\pm\bm{n}$ because for the magnetic moment $\bm{e}$ the directions $\bm{n}$
and $\bm{-n}$ are equivalent. At zero temperature, the magnetic moment $\bm{e}$, once located
in a particular potential well, is confined there forever. At finite temperature, the
probability of an overbarrier (interwell) transition becomes non-zero. If the ratio
$\sigma\equiv Kv/kT$ is high enough, the transition rate is exponential thus yielding the
N\'eel law $\tau\propto\exp(\sigma)$ for the reference time $\tau$ of the particle
re-magnetization.

Brown\cite{Brown-PR:63} shaped up those semi-qualitative considerations into a rigorous
Sturm-Liouville eigenvalue problem by deriving the micromagnetic kinetic equation
\begin{equation}                                                        \label{eq:02}
2\tau_D\,\partial W/\partial t=\widehat{\bm{J}}W\widehat{\bm{J}}(U/kT+\ln W),
\end{equation}
where $W(\bm{e},t)$ is the orientational distribution function of the magnetic moment,
$\widehat{\bm{J}}=(\bm{e}\times\partial/\partial \bm{e})$ is the infinitesimal rotation
operator with respect to $\bm{e}$, and the time $\tau_D$ is introduced below by formula
(\ref{eq:04}). Generally speaking, equation~(\ref{eq:02}) is incomplete since a
gyromagnetic term is absent there. This means that the consideration is limited by the
frequency range $\omega\tau_0\ll1$, where $\tau_0$ is the relaxation time of the Larmor
precession of the particle magnetic moment in the internal anisotropy field $H_a\sim
=2K/I$, where $K$ includes the possible shape contribution. Comparing this condition with
the other one, $\omega_L\tau_0\lesssim1$, that evidences a low-to-moderate quality factor
of the Larmor precession for real nanodisperse ferrites, one estimates the allowed
frequency as $\omega\ll\omega_L$ that means, in fact, a fairly wide range.\footnote{We
remark that, in principle, there might occur a situation, where $\tau$ is very long, i.e.,
the quality factor of precession is very high. Then the gyromagnetic term in the kinetic
equation must be retained so that Larmor precession begins to interact with the
superparamagnetic (longitudinal) relaxation. An example of such a situation is considered
in Ref.~\protect\onlinecite{GaSv-PRL:00}, where some interesting nonlinear effects are
found.}

In the statistical description delivered by Eq.~(\ref{eq:02}), the observed (macroscopic)
magnetic moment per particle is given by the average
\begin{equation}                                                        \label{eq:03}
\bm{m}(t)=\mu\langle\,\bm{e}\,\rangle=\int\,\bm{e}\,W(\bm{e},t)\,d\bm{e},
\end{equation}
Note that with allowance for Eq.~(\ref{eq:01}) the function $W$ has a parametric dependency on
the vector $\bm{n}$ so that, in fact, the angular argument of $W$ is $(\bm{e}\cdot\bm{n})$.

The magnetodynamic equation underlying the Brown kinetic equation~(\ref{eq:02}) can be either
that by Landau \&{} Lifshitz or that by Gilbert. To be specific, we adopt the former one.
Thence, the reference relaxation time in Eq.~(\ref{eq:02}) writes as
\begin{equation}                                                        \label{eq:04}
\tau_D=Iv/2\alpha\gamma kT,
\end{equation}
where $\gamma$ is the gyromagnetic ratio for electrons and $\alpha$ is the precession damping
(spin-lattice relaxation) phenomenological parameter.

Assuming uniaxial symmetry of the time-dependent solution and separating the variables in
Eq.~(\ref{eq:02}) in the form
\begin{equation}                                                        \label{eq:05}
W(\bm{e},t)=\textstyle{\frac{1}{2\pi}}\sum_{\ell=0}^{\infty}\,A_\ell\,\psi_\ell(\bm{e}
\cdot\bm{n})\,\exp(-\lambda_\ell t/2\tau_D),
\end{equation}
where the amplitudes $A_\ell$ depend on the initial perturbation, one arrives at the spectral
problem
\begin{equation}                                                        \label{eq:06}
\widehat{L}\psi_\ell=\lambda_\ell\psi_\ell, \qquad
\widehat{L}\equiv\widehat{\bm{J}}\left[\,2\sigma(\bm{e}\cdot\bm{n})
(\bm{e}\times\bm{n})-\widehat{\bm{J}}\,\right],
\end{equation}
where the non-negativity of the decrements $\lambda_\ell$ can be proven easily. Expanding the
eigenmodes $\psi_\ell$ in the Legendre polynomial series
\begin{equation}                                                        \label{eq:07}
\psi_\ell=\textstyle{\frac{1}{2}}\sum\limits_{k=1}^{\infty}(2k+1)\,b_k^{(\ell)}
P_k(\cos\theta), \quad k=\,1,\, 3,\, 5,\,\ldots\;,
\end{equation}
where $\theta$ is the angle between $\bm{e}$ and $\bm{n}$, one arrives at the homogeneous
tridiagonal recurrence relation
\begin{widetext}
\begin{equation}                                                        \label{eq:08}
\left[1-\frac{\lambda_\ell}{k(k+1)}\right]\;b_k^{(\ell)}
-2\sigma\left[\frac{k-1}{(2k-1)(2k+1)}\;b_{k-2}^{(\ell)}+ \frac{1}{(2k-1)(2k+3)}\>
b_k^{(\ell)}-\frac{k+2}{(2k+1)(2k+3)}\; b_{k+2}^{(\ell)}\right]=0,
\end{equation}
\end{widetext}
Note that Eqs.~(\ref{eq:05})--(\ref{eq:08}) describe only the longitudinal (with respect to
the easy axis) relaxation of the magnetic moment. We remark that under condition
$\omega\ll\omega_{\rm L}$, i.e., far from the ferromagnetic resonance range, the transversal
components of $\bm{m}=\mu\langle\,\bm{e}\,\rangle$ are of minor importance.

\subsection{Interwell mode \label{sec:2.2}}
Spectral equation~(\ref{eq:06}) describes the temperature-induced (fluctuation) motions of the
vector $\bm{e}$ in the orientational potential with a symmetrical profile~(\ref{eq:01}). With
respect to the time dependence, the set of possible eigenmodes splits into two categories:
interwell (overbarrier) transitions and intrawell wanderings. In the spectral
problem~(\ref{eq:06}) the interwell transitions of the magnetic moment are associated with the
single eigenvalue $\lambda_1$. As the rigorous analysis shows,\cite{Stor-K:85} it drastically
differs from the others: whereas for $\ell\geq1$ all the $\lambda_\ell$ gradually {\em grow}
with $\sigma$, the decrement $\lambda_1$ exponentially {\em falls down} proportionally to
$\exp(-\sigma)$.

In the opposite limit $\sigma\rightarrow0$, all the decrements, including $\lambda_1$, tend to
the sequence $\lambda_\ell=\ell(\ell+1)$ and thus become of the same order of magnitude. This
regime corresponds to a vanishing anisotropy so that the difference between the inter- and
intrawell motions disappear, and the magnetic moment diffuses almost freely over all the
$4\pi$ radians with the reference time $\tau_D$ introduced by Eq.~(\ref{eq:04}).

From Eqs.~(\ref{eq:03}) and (\ref{eq:05}) one finds that the longitudinal component of the
magnetic moment evolves according to
\begin{equation}                                                        \label{eq:09}
m(t)=\mu\,\sum_{\ell=1}^{\infty}\,A_\ell\,e^{-\lambda_\ell t/2\tau_D}
\int_{-1}^{1}x\psi_\ell\,dx,
\end{equation}
where $x=\cos\theta=(\bm{e}\cdot\bm{n})$. For a symmetrical potential like~(\ref{eq:01}) the
equilibrium value $m_0$ of the particle magnetic moment is zero.

With the above-mentioned structure of the eigenvalue spectrum, the term with $\ell=1$ in
Eq.~(\ref{eq:09}), being proportional to $\exp(-e^{-\sigma}t/\tau_D)$, at $\sigma>1$ is far
more long-living than any other one. The dominating r\^ole of the decrement $\lambda_1$ had
been proven by Brown, and for it he had derived\cite{Brown-PR:63} the asymptotic expression
\begin{equation}                                                        \label{eq:10}
\lambda_{\text{B}}=(4/\sqrt{\pi})\,\sigma^{3/2}\,e^{-\sigma}, \qquad (\sigma\gg1).
\end{equation}
In a short time after, using a continued fraction method, Aharoni
constructed\cite{Ahar-PR:64} for $\lambda_1$ a fairly long power series in $\sigma$ and
also showed numerically that Brown's expression~(\ref{eq:10}) resembles the exact one with
the accuracy of several percent for $\sigma\gtrsim3$. In the 90's the eigenvalue
$\lambda_1$ became a subject of extensive studies. Efficient numerical procedures were
developed \cite{Kalm-PRE:00} and a number of extrapolation formulas with a good overall
accuracy were proposed\cite{BeBe-PRB:92,Ahar-PRB:92,CoCr-JMMM:94,CrCr-JAP:94}.

\subsection{Asymptotic solution of the Brown equation \label{sec:2.3}}
The study that we describe below was inspired by our work on fitting the dynamic
susceptibilities measurements for real assemblies of fine particles. Those data typically
describe polydisperse systems in the low-frequency bandwidth $\omega/2\pi=1$--10$^3$\,Hz. As
$\tau_0\sim10^{-9}$\,s or smaller, then, using formula~(\ref{eq:10}) for estimations, one
concludes that the mentioned frequency interval becomes a dispersion range for the interwell
(superparamagnetic) mode at
\begin{displaymath}
\omega\tau_0e^\sigma\gtrsim1 \quad {\rm that\>\ is\ } \quad \sigma\gtrsim10.
\end{displaymath}
For temperatures up to 300\,K this condition holds for quite a number of nanomagnetic systems.

Application of the best fit procedure to a set of experimental data implies numerous
re-calculations of the linear and non-linear susceptibility curves $\chi^{(k)}$ of the
assembly. Any such curve, due to a considerable polydispersity of the particles, is a
superposition of a great number of partial curves $\chi^{(k)}(\sigma)$ spread over a wide size
(or, in the dimensionless form, $\sigma$) range. For successful processing, one needs a fast
and very accurate algorithm to evaluate $\chi^{(k)}(\sigma)$ everywhere including the domain
$\sigma\gg1$. The existing extrapolation formulas are no good for that purpose due to their
ill-controllable error accumulation. A plausible way out is an asymptotic in $\sigma^{-1}$
solution of Eq.~(\ref{eq:06}). In the course of the fitting procedure, this approximation can
be easily matched in the intermediate $\sigma$ range with the well-known expansions for the
small $\sigma$ end.

It is noteworthy that some 20 years ago Brown himself resumed\cite{Brown-P:77,Brown-IEEE:79}
studies on $\lambda_1$ and modified the pre-exponential factor in Eq.~(\ref{eq:10})
transforming it into an asymptotic series in $\sigma^{-1}$. On the base of Eq.~(\ref{eq:06})
he had constructed an integral recurrence procedure, and evaluated $\lambda_1$ down to terms
$\propto1/\sigma^{10}$. What we do below, is, in fact, carrying on this line of analysis that
had not been touched since then. Our method advances Brown's results in two aspects. First,
for $\lambda_1$ it is more simple. Second, it provides not only the eigenvalue but the
eigenfunction as well. Only having the latter in possession, one is able to obtain theoretical
expressions for the directly measurable quantities that is the susceptibilities $\chi^{(k)}$.

Taking Eq.~(\ref{eq:06}) as the starting point, we remark its equilibrium solution
\begin{eqnarray}                                                        \label{eq:11}
\psi_0=Z_0^{-1}\exp(\sigma x^2), \quad Z_0=2R(\sigma),                  \\
R(\sigma)=\int_{0}^{1}\,\exp(\sigma x^2)\,dx,                    \nonumber
\end{eqnarray}
that corresponds to $\ell=0$ and $\lambda_0=0$ and note the asymptotic expansion for the
partition integral $R(\sigma)$ found in Ref.~\onlinecite{RaSh-JETP:74}:
\begin{widetext}
\begin{equation}                                                        \label{eq:12}
R(\sigma)=e^\sigma G/2\sigma, \quad G(\sigma)\equiv1+\frac{1}{2\sigma} +\frac{3}{4\sigma^2}+
\frac{15}{8\sigma^3}+\ldots+\frac{(2n-1)!!}{2^n\sigma^n}+\ldots.
\end{equation}
\end{widetext}

The operator $\widehat L$ in Eq.~(\ref{eq:06}) is not self-conjugated and thus produces two
sets of eigenfunctions, which obey the respective equations
\begin{equation}                                                        \label{eq:13}
\widehat L\psi_k=\lambda_k\psi_k\,, \qquad \widehat L^{+}\varphi_j=\lambda_j\varphi_j\,;
\end{equation}
here $+$ denotes Hermitian conjugation. The eigenfunctions of these two families are
orthonormalized and related to each other in a simple way:
\begin{equation}                                                        \label{eq:14}
\psi_k=\psi_0\varphi_k\,, \qquad \int_{-1}^1dx\,\varphi_j\psi_k=\delta_{jk}\,.
\end{equation}
Qualitatively, from Eq.~(\ref{eq:14}) one may say that $\varphi_k$ are the same eigenfunctions
but ``stripped'' of the exponential equilibrium solution $\psi_0$. Substituting
Eq.~(\ref{eq:14}) in Eq.~(\ref{eq:06}), one gets two useful relationships
\begin{equation}                                                        \label{eq:15}
-\widehat{\bm{J}}\psi_0\widehat{\bm{J}}\varphi_k =\lambda_k\psi_0\varphi_k, \qquad \int\psi_0
(\widehat{\bm{J}}\varphi_j)(\widehat{\bm{J}}\varphi_k)\,dx =\lambda_k\delta_{jk}\,,
\end{equation}
where the second one follows from the first after multiplication by $\varphi_i$ and
integration by parts. Note that in the second formula action of each operator reaches no
farther than the nearest closing parenthesis.

On rewriting Eq.~(\ref{eq:15}.1) in terms of a single orientational variable
$x=(\bm{e}\cdot\bm{n})$, the spectral problem takes the form
\begin{equation}                                                        \label{eq:16}
\frac{d}{dx}\,\left[\,\psi_0(1-x^2)\,\frac{d\varphi_k}{dx}\right] =-\lambda_k\psi_0\varphi_k.
\end{equation}
In the equilibrium state Eq.~(\ref{eq:16}) reduces to
\begin{equation}                                                        \label{eq:17}
\frac{d}{dx}\,\left[\,\psi_0(1-x^2)\,\frac{d\varphi_0}{dx}\right]=0,
\end{equation}
whose normalized solution is $\varphi_0=1$. This solution, being a true equilibrium one, turns
the inner part of the brackets, i.e., the probability flux in the kinetic equation
(\ref{eq:02}), into identical zero.

As remarked in Sec.~\ref{sec:2.2}, at $\sigma\gg1$ the most long-living non-stationary
solution of Eq.~(\ref{eq:16}) is the eigenfunction with $\ell=1$, whose eigenvalue is
exponentially small, see Brown's estimation~(\ref{eq:10}). We use this circumstance for
approximate evaluation of $\varphi_1$ in the $\sigma\gg1$ limit by neglecting the right-hand
side of Eq.~(\ref{eq:16}) for $\ell=1$. On doing that, the equation obtained for the function
$\varphi_1$ formally coincides with equation~(\ref{eq:17}) for $\varphi_0$. However, the
essential difference is that now the content of the bracket is non-zero:
\begin{equation}                                                        \label{eq:18}
\psi_0(1-x^2)\,\frac{d\varphi_1}{dx}=\textstyle{\frac{1}{2}}C,
\end{equation}
where $\frac{1}{2}C$ is the integration constant. Note also that, contrary to $\varphi_0$, the
sought for solution $\varphi_1$ is odd in $x$.

\begin{figure}
\includegraphics[width=5.5cm]{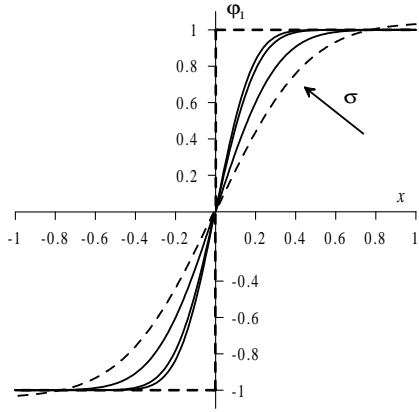}
\caption{Eigenmode $\varphi_1(x)$ determined with the aid of the numerical solution of
Eq.~(\protect\ref{eq:08}) for the dimensionless barrier height $\sigma$: 5 (dashed line), 10,
20, 25 (solid lines); the arrow shows the direction of $\sigma$ growth. Thick dashes show the
step-wise function that is the limiting contour for $\varphi_1$ at $\sigma\rightarrow\infty$.
\label{fig:01}}
\end{figure}

Using the explicit form of $\psi_0$ from Eq.~(\ref{eq:11}) and integrating, one gets for $x>0$
\begin{widetext}
\begin{equation}                                                        \label{eq:19}
\varphi_1=CR\int_{0}^{x}\frac{e^{-\sigma x^2}}{1-x^2}\,dx
=CR\int_{0}^{x}\,e^{-\sigma x^2}\,(1+x^2+x^4+x^6+\ldots)\,dx.
\end{equation}
\end{widetext}
The integrals in expansion~(\ref{eq:19}) are akin. Denoting
\begin{displaymath}
F_n=\int_0^x\,x^{2n}\,e^{-\sigma x^2}\,dx,
\end{displaymath}
one can easily write for them the recurrence relation and ``initial'' condition as
\begin{equation}                                                        \label{eq:20}
F_n=-\frac{\partial}{\partial\sigma}F_{n-1}\,,  \qquad
F_0=\frac{\sqrt{\pi}}{2\sqrt{\sigma}}\,\,{\rm erf}(\sqrt{\sigma}x),
\end{equation}
respectively. Using the asymptotics of the error integral, with the exponential accuracy in
$\sigma$ one finds
\begin{equation}                                                        \label{eq:21}
F_n=[(2n-1)!!/2^n\sigma^n]\,F_0\,, \qquad F_0\simeq\sqrt{\pi}/2\sqrt{\sigma}.
\end{equation}
Comparing this with expression~(\ref{eq:12}) for the function $G$, we get the representation
\begin{equation}                                                        \label{eq:22}
\varphi_1(x>0)\simeq CRF_0G.
\end{equation}
Applying to Eq.~(\ref{eq:22}) the normalizing condition~(\ref{eq:14}), one evaluates the
constant as $C=1/RF_0G$. Therefore, from Eqs.~(\ref{eq:20})--(\ref{eq:22}) the principal
relaxational eigenmode determined with the $\exp(-\sigma)$ accuracy emerges as an odd step
function
\begin{equation}                                                        \label{eq:23}
\varphi_1(x)\simeq\begin{cases}-1 & \text{for}\>\, x<0,\\ 1 & \text{for}\>\, x>0.\end{cases}
\end{equation}
In Fig.~\ref{fig:01} the limiting contour~(\ref{eq:23}) is shown against the exact curves
$\varphi_1(x)$ obtained by solving numerically Eq.~(\ref{eq:08}) for several values of
$\sigma$. We remark that in the statistical calculations carried out below the typical
integrals are of two kinds. In the first, the integrand consists of the product of
$\varphi_1\psi_0$ and some non-exponential function. As $\psi_0\propto\exp{\sigma x^2}$, the
details of behavior of $\varphi_1$ in the vicinity of $x=0$ are irrelevant because the
approximate integral will differ but exponentially from the exact result. The integrals of the
second type contain $d\varphi_1/dx$ in the integrand. For them a step-wise approximation
Eq.~(\ref{eq:23}) with its derivative equal identical zero everywhere except for $x=0$ is an
inadmissible choice. So, to keep the exponential accuracy in this case, one has to get back to
Eq.~(\ref{eq:18}).

\begin{figure}
\includegraphics[width=5.5cm]{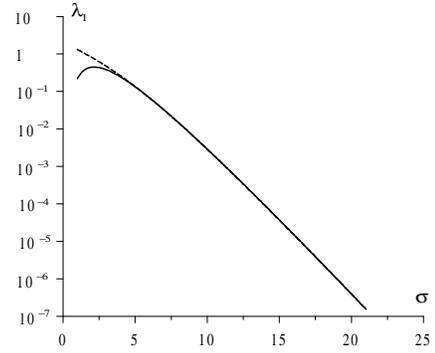}
\caption{Asymptotic expression (\protect\ref{eq:25}) for the eigenvalue $\lambda_1$ with
allowance for terms up to $\sigma^{-9}$ (solid line) compared to the exact numeric value
(dashed line). \label{fig:02}}
\end{figure}

The eigenvalue $\lambda_1$ corresponding to the approximate eigenfunction $\varphi_1$ from
Eq.~(\ref{eq:23}) is evaluated via formula~(\ref{eq:15}) that can be rewritten as
\begin{equation}                                                        \label{eq:24}
\lambda_1=\int_{-1}^{1}\,\psi_0\,\left(\widehat{\bm{J}}\varphi_1\right)^2=
\frac{1}{R}\int_0^1\,e^{\sigma x^2}\,(1-x^2)\,\left(\frac{d\varphi_1}{dx} \right)^2\,dx.
\end{equation}
Substituting the derivative from Eq.~(\ref{eq:18}), one finds
\begin{displaymath}
\lambda_1=C=(2/\sqrt{\pi})\,\,\sigma^{1/2}/RG,
\end{displaymath}
and using expression~(\ref{eq:12}) for $R$ finally arrives at
\begin{equation}                                                        \label{eq:25}
\lambda_1=(4/\sqrt{\pi})\,\,\sigma^{3/2}e^{-\sigma}/G^2=\lambda_{\text{B}}/G^2.
\end{equation}
With $G$ expanded in powers of $\sigma^{-1}$, see Eq.~(\ref{eq:12}), this formula reproduces
the asymptotic expression derived by Brown in Ref.~\onlinecite{Brown-P:77}. At $G=1$ it
reduces to his initial result,\cite{Brown-PR:63} corresponding to the above-given
Eq.~(\ref{eq:10}). Function $\lambda_1(\sigma)$ from Eq.~(\ref{eq:25}) is shown in
Fig.~\ref{fig:02} in comparison with the exact result obtained by a numerical solution.
Indeed, at $\sigma\gtrsim3$ the results virtually coincide.

According to expansion~(\ref{eq:05}), each decrement $\lambda_\ell$ defines the reference
relaxation time
\begin{equation}                                                        \label{eq:26}
\tau_\ell=2\tau_D/\lambda_\ell.
\end{equation}
Thence from Eq.~(\ref{eq:25}) we get
\begin{equation}                                                        \label{eq:27}
\tau_1=2\tau_D/\lambda_1=\tau_{\text{B}}\,G^2, \qquad
\tau_{\text{B}}\equiv2\tau_D/\lambda_{\text{B}},
\end{equation}
where $\tau_{\text{B}}$ denotes the asymptotic relaxation time obtained by Brown in
Ref.~\onlinecite{Brown-PR:63}. Substituting in Eq.~(\ref{eq:27}) the explicit asymptotic
series~(\ref{eq:12}) for $G$, one gets
\begin{equation}                                                        \label{eq:28}
\tau_1=\tau_D\frac{\sqrt{\pi}e^\sigma}{2\sigma^{3/2}}\,
\left(1+\frac{1}{\sigma}+\frac{7}{4\sigma^2}+\frac{9}{2\sigma^3}+\ldots \right).
\end{equation}

\subsection{Asymptotic integral time \label{sec:2.4}}
The decrements $\lambda_\ell$ or, equivalently, relaxation times $\tau_\ell$, being the
characteristics of the eigenfunctions of the distribution function, are not observable if
taken as separate quantities. However, in combination they are involved in a useful directly
measurable quantity, the so-called integral relaxation time. In terms of correlation functions
this characteristics is defined as
\begin{equation}                                                        \label{eq:29}
\tau_{\rm int}= \int_0^{\infty}\frac{\langle\,m(t)m(0)\,\rangle_0}
{\langle\,m^2(0)\,\rangle_0}\,dt=\int_0^{\infty}\frac{\langle\,x(t)x(0)
\,\rangle_0}{\langle\,x^2(0)\,\rangle_0}\,dt,
\end{equation}
where the angular brackets stand for the statistical ensemble averaging over the equilibrium
distribution~(\ref{eq:12}). As follows from Eq.~(\ref{eq:29}), the integral relaxation time
equals the area under the normalized decay of magnetization.

The Green function of Eq.~(\ref{eq:02}), i.e., the probability density of a state $(x,t)$,
provided the initial state is $(x_0,0)$, writes
\begin{equation}                                                        \label{eq:30}
W(x,t;x_0,0)=\sum_{\ell=0}^{\infty}\,\psi_\ell(x)\,\varphi_\ell(x_0)\,e^ {-\lambda_\ell t}.
\end{equation}
Similarly to Eq.~(\ref{eq:07}), we expand the eigenfunctions in Legendre polynomials as
\begin{equation}                                                        \label{eq:31}
\psi_\ell=\textstyle{\frac{1}{2}}\sum\limits_{k=1}^{\infty}(2k+1)\,b_k^{(\ell)}P_k(x), \qquad
\varphi_\ell=\sum\limits_{k=1}^{\infty}\,a_k^{(\ell)}P_k(x).
\end{equation}
and introduce special notations for the first two functions
\begin{eqnarray}                                                        \label{eq:32}
&& \psi_0=\textstyle{\frac{1}{2}}\sum\limits_{k=0}^{\infty}(2k+1)\,S_kP_k(x), \\
&& \psi_1=\textstyle{\frac{1}{2}}\sum\limits_{k=0}^{\infty}(2k+1)\,Q_kP_k(x). \nonumber
\end{eqnarray}
The procedures to evaluate the coefficients $S_k$ and $Q_k$ and the explicit asymptotic forms
for $Q_1$ and $S_2$ are given in Appendix \ref{sec:AppA}; note representation~(\ref{eq:11})
for the equilibrium function $\psi_0$.

Due to Eq.~(\ref{eq:14}), the coefficients in formulas~(\ref{eq:31}) are related to each other
by $b_k^{(\ell)}=\langle P_kP_{k'}\rangle_0\,a_{k'}^{(\ell)}$. In those terms one gets for the
correlator in Eq.~(\ref{eq:14}):
\begin{widetext}
\begin{equation}                                                        \label{eq:33}
\langle\langle x(t)x(0)\rangle\!\rangle_0= \int\int xx_0\,\psi_0\,W(x,t;x_0,0)\,dx\,dx_0
=\sum_{\ell=1} \left[b_1^{(\ell)}\right]^2\,e^{-\lambda_\ell t/2\tau_D},
\end{equation}
\end{widetext}
where the averaging over the current coordinate $x$ is performed with the function $W$ from
Eq.~(\ref{eq:30}) whereas that over the initial conditions---with the equilibrium function
$\psi_0$. Substituting expression~(\ref{eq:33}) in Eq.~(\ref{eq:29}) one gets the integral
time in the form
\begin{equation}                                                        \label{eq:34}
\tau_{\rm int}=\sum_{\ell=1}^{\infty}\,\tau_\ell\left[b_1^{(\ell)}\right]^2
\big/\sum_{\ell=1}^{\infty}\left[b_1^{(\ell)}\right]^2
=\sum_{\ell=1}^{\infty}\,\tau_\ell\left[b_1^{(\ell)}\right]^2 \big/\langle\,x^2\,\rangle_0\,.
\end{equation}

Unlike $\tau_1$, which in principle cannot be evaluated\cite{Coff-ACP:98} analytically at
arbitrary $\sigma$, for $\tau_{\text{int}}$ an exact solution is possible for arbitrary values
of the anisotropy parameter. Recently two ways were proposed to obtain quadrature formulas for
$\tau_{\text{int}}$. One method\cite{GaIs-TMF:90} implies a direct integration of the
Fokker-Planck equation. Another method\cite{CoCr-PRE:94} involves solving three-term
recurrence relations for the statistical moments of $W$. The emerging solution for
$\tau_{\text{int}}$ can be expressed in a finite form in terms of hypergeometric (Kummer's)
functions. Equivalence of both approaches was proven in Ref.~\onlinecite{CoCr-PRE:96}.

In the present study, as mentioned, we are dealing in the high-barrier approximation. In this
limiting case $\lambda_1$ is exponentially small, so that the term with $\ell=1$ in the
numerator in Eq.~(\ref{eq:34}) is far greater than the others. With allowance for
Eq.~(\ref{eq:32}) it can be written as
\begin{equation}                                                        \label{eq:35}
\tau_{\rm int}=\tau_1\,\left[b_1^{(1)}\right]^2\big/\langle\,x^2\,\rangle_0
=\tau_1\,Q_1^2\big/\langle\,x^2\,\rangle_0.
\end{equation}
The equilibrium moment calculated by definition writes
\begin{equation}                                                        \label{eq:36}
\langle\,x^2\,\rangle_0=(1/2\sigma)\,(e^\sigma-1)=1/G-1/2\sigma,
\end{equation}
and for $\sigma\gg1$, using formula~(\ref{eq:A05}) of Appendix~\ref{sec:AppA} we get
\begin{equation}                                                        \label{eq:37}
Q_1\simeq 1/G.
\end{equation}
Substitution of Eqs.~(\ref{eq:36}) and (\ref{eq:37}) in (\ref{eq:35}) with allowance for
relationships~(\ref{eq:12}), (\ref{eq:25}) and (\ref{eq:27}) gives the asymptotic
representation in the form
\begin{eqnarray}                                                        \label{eq:38}
\tau_{\rm int} & = & \tau_{\text{B}}\,\frac{2\sigma G}{(2\sigma-G)}           \\
& = & \tau_D\frac{\sqrt{\pi}e^\sigma}
{2\sigma^{3/2}}\,\left(1+\frac{1}{\sigma}+\frac{3}{2\sigma^2}+\frac{13}{4\sigma^3}+\ldots
\right).                                                                \nonumber
\end{eqnarray}
As it is seen from formulas~(\ref{eq:28}) and (\ref{eq:38}) written with the accuracy up to
$\sigma^{-3}$, the asymptotic expressions for the interwell and integral times deviate
beginning with the term $\propto\sigma^{-2}$. This contradicts the only known to us asymptotic
expansion of $\tau_{\text{int}}$ given in Eq.~(60) of Refs.~\onlinecite{CoCr-PRE:94} and
repeated in Eq.~(7.4.3.22) of the book \cite{CoKa-TLE:96}. The latter expression written with
the accuracy $\propto\sigma^{-2}$, instead of turning into Eq.~(\ref{eq:38}) coincides with
the Brown's expression~(\ref{eq:28}) for $\tau_1$. Meanwhile, as it follows from formula
(\ref{eq:35}), such a coincidence is impossible and therefore Eqs.~(60) of
Refs.~\onlinecite{CoCr-PRE:94} and (7.4.3.22) of Ref.~\onlinecite{CoKa-TLE:96} are misleading.
The necessity to rectify this issue made us to begin the demonstration of our approach with
the case of the integral relaxation time. Further on we consistently apply our procedure for
description of the nonlinear (third- and fifth-harmonic) dynamic susceptibilities of a solid
superparamagnetic dispersion.

\section{Perturbative expansions for the distribution function \label{sec:3}}
\subsection{Static probing field \label{sec:3.1}}
To find the nonlinear susceptibilities, one has to take into account the changes that the
probing field induces in the basic state of the system. In the limit $\sigma\gg1$, which we
deal in, the relaxation time $\tau_1$ of the interwell mode $\psi_1$ is far greater than all
the other relaxation times $\tau_k$. This means that with respect to the intrawell modes the
distribution function is in equilibrium. So it suffices to determine the effect of the probing
field $\bm{H}=H\bm{h}$ just on $\psi_0$ and $\psi_1$. Assuming the energy function in the
form
\begin{equation}                                                        \label{eq:39}
U+U_H=-Kv(\bm{e}\cdot\bm{n})^2-IvH(\bm{e}\cdot\bm{h}),
\end{equation}
[compare with Eq.~(\ref{eq:01})], and separating variables in Eq.~(\ref{eq:02}), one arrives
at the eigenfunction problem
\begin{equation}                                                        \label{eq:40}
\widehat{L}\,f_\beta=\xi\widehat{V}\,f_\beta,
\end{equation}
where $\xi=IvH/kT$ and notation $f_\beta$ refers to the distribution function modes that stem
from $\psi_0$ or $\psi_1$ at $H\neq0$, i.e., $\beta=0$ or 1. In Eq.~(\ref{eq:40}) operator
$\widehat{L}$ is defined by Eq.~(\ref{eq:06}) whilst
$\widehat{V}=-\xi\widehat{\bm{J}}(\bm{e}\times\bm{h})$ is the operator caused by the energy
term $U_H$ in (Eq.~\ref{eq:39}). As in above, for the non-self-conjugated spectral problem
(\ref{eq:40}) we introduce the family of conjugated functions $g_\beta$ and set
$f_\beta=g_\beta \psi_0$.

Following our approach, in the low-temperature limit ($\sigma\gg1$) we \textit{set to zero}
the eigenvalues corresponding to both $f_0$ and $f_1$; compare with Eqs.~(\ref{eq:17}) and
(\ref{eq:18}) for $\psi_0$ and $\psi_1$. Assuming the temperature-scaled magnetic field $\xi$
to be small, we treat $U_H$ as a perturbation Hamiltonian and expand the principal
eigenfunctions as
\begin{equation}                                                        \label{eq:41}
f_0=\sum_{n=0}\,\xi^n f_0^{(n)}, \qquad f_1=\sum_{n=0}\,\xi^n f_1^{(n)}.
\end{equation}
Thence for the field-free ($H=0$) case one has $f_0^{(0)}=\psi_0$ and $f_1^{(0)}=\psi_1$. The
same kind of expansion is assumed for $g_\beta$ with $g_0^{(0)}=1$ and $g_1^{(0)}=\varphi_1$.
Note also that in order to retain the normalizing condition we require that $f_\beta^{(n)}$
have zero averages.

Substituting expansion~(\ref{eq:41}) in Eq.~(\ref{eq:40}) and collecting the terms of the same
order in $\xi$, we arrive at the recurrence relation
\begin{equation}                                                        \label{eq:42}
\widehat{L}\,f_\beta^{(n)}=\widehat{V}\,f_\beta^{(n-1)},
\end{equation}
that for the particular cases $\beta=0$ and 1 with the aid of the identity
$\bm{e}\times\bm{h}=\widehat{\bm{J}}(\bm{e}\cdot\bm{h})$ takes the formes
\begin{eqnarray}                                                        \label{eq:43}
&& \widehat{\bm{J}}\psi_{0}\widehat{\bm{J}}g_{0}^{(n)}=
\widehat{\bm{J}}\psi_{0}g_{0}^{(n-1)}\widehat{\bm{J}}(\bm{e}\cdot\bm{h}), \\
&& \widehat{\bm{J}}\psi_{0}\widehat{\bm{J}}g_{1}^{(n)}=
\widehat{\bm{J}}\psi_{0}g_{1}^{(n-1)}\widehat{\bm{J}}(\bm{e}\cdot\bm{h}), \nonumber
\end{eqnarray}
respectively. Set~(\ref{eq:43}) solves easily for $g_0$ since $g_0^{(0)}=\varphi_0=1$.
Starting with $n=0$, one gets sequentially
\begin{widetext}
\begin{eqnarray}                                                        \label{eq:44}
g_{0}^{(1)}&=&(\bm{e}\cdot\bm{h}),
                                                                        \\
g_{0}^{(2)}&=&\textstyle{\frac{1}{2}}\left[(\bm{eh})^2-\langle\,(\bm{e}\cdot
\bm{h})^2\,\rangle_0\right],
                                                                        \nonumber \\
g_{0}^{(3)}&=&\textstyle{\frac{1}{6}}(\bm{e}\cdot\bm{h})^3-\textstyle{\frac{1}{2}}(\bm{e}
\cdot\bm{h})\,\langle\,(\bm{e}\cdot\bm{h})^2\,\rangle_0,
                                                                        \nonumber\\
g_{0}^{(4)}&=&\textstyle{\frac{1}{24}}\left[(\bm{e}\cdot\bm{h})^4-\langle\,
(\bm{e}\cdot\bm{h})^4\,\rangle_0\right]-
\textstyle{\frac{1}{4}}\left[(\bm{e}\cdot\bm{h})^2\,\langle\,(\bm{e}\cdot\bm{h})
^2\,\rangle_0-\langle\,(\bm{e}\cdot\bm{h})^2\,\rangle_0^2\right],
                                                                        \nonumber \\
g_{0}^{(5)}&=&\textstyle{\frac{1}{120}}(\bm{e}\cdot\bm{h})^5-\textstyle{\frac{1}{12}}
(\bm{e}\cdot\bm{h})^3\,\langle\,(\bm{e}\cdot\bm{h})^2\,\rangle_0
-\textstyle{\frac{1}{24}}(\bm{e}\cdot\bm{h})\left[\langle\,(\bm{e}\cdot\bm{h})
^4\,\rangle_0-6\langle\,(\bm{e}\cdot\bm{h})^2\,\rangle_0^2\right]       \nonumber.
\end{eqnarray}
All the obtained functions are constructed in such a way that the corresponding
$f_\beta^{(n)}$ satisfy the above-mentioned zero average requirement. We remark also that
there is no problem to continue the calculational procedure to any order.

Evaluation of $g_1$ is done in two steps. At the first one, we set $g_1^{(0)}$ equal to the
antisymmetric step-wise function~(\ref{eq:23}) and its derivative equal zero. After that from
the second of Eqs.~(\ref{eq:43}) we can express $g_{1}^{(k)}$ in closed form. Taken up to the
fourth order these ``zero-derivative'' solutions write
\begin{eqnarray}                                                        \label{eq:45}
g_{1}^{(1)}&=&\varphi_1\,(\bm{e}\cdot\bm{h})-\langle\,
\varphi_1(\bm{e}\cdot\bm{h})\,\rangle_0,                         \\
g_{1}^{(2)}&=&\textstyle{\frac{1}{2}}\,\varphi_1(\bm{e}\cdot\bm{h})^2-
(\bm{e}\cdot\bm{h})\,\langle\,\varphi_1(\bm{e}\cdot\bm{h})\, \rangle_0,
                                                                        \nonumber\\
g_{1}^{(3)}&=&\textstyle{\frac{1}{6}}\left[\varphi_1\,(\bm{e}\cdot
\bm{h})^3-\langle\,\varphi_1(\bm{e}\cdot\bm{h})^3\,\rangle_0
\right]-\textstyle{\frac{1}{2}}\langle\,\varphi_1(\bm{e}\cdot\bm{h})\,
\rangle_0\,\left[(\bm{e}\cdot\bm{h})^2-\langle\,(\bm{e}\cdot \bm{h})^2\,\rangle_0\right],
                                                                        \nonumber\\
g_{1}^{(4)}&=&\textstyle{\frac{1}{24}}\varphi_1\,(\bm{e}\cdot\bm{h})^4
-\textstyle{\frac{1}{6}}\langle\,\varphi_1\,(\bm{e}\cdot
\bm{h})\,\rangle_0\,\left[(\bm{e}\cdot\bm{h})^3-3(\bm{e}
\cdot\bm{h})\,\langle\,(\bm{e}\cdot\bm{h})^2\,\rangle_0\right]
                                                                        \nonumber\\
& &\hspace*{88mm}
-\textstyle{\frac{1}{6}}(\bm{e}\cdot\bm{h})\,\langle\,\varphi_1\,(\bm{e}\cdot
\bm{h})^3\,\rangle_0\,                                                  \nonumber.
\end{eqnarray}
Note the alternating parity in $\bm{e}$ with the term order growth in both Eqs.~(\ref{eq:44})
and (\ref{eq:45}).
\end{widetext}

\begin{figure}
\includegraphics[width=5.5cm]{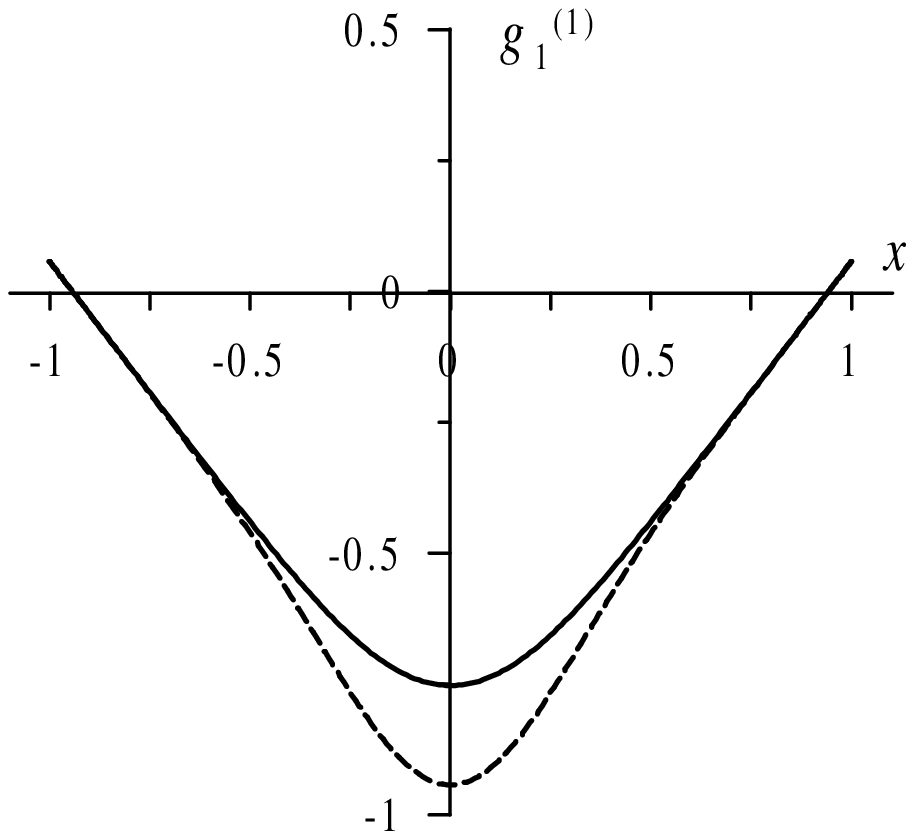}
\caption{Function $g_1^{(1)}$ found numerically (solid) and evaluated in the
``zero-derivative'' approximation (dashed). \label{fig:03}}
\end{figure}

\begin{figure}
\includegraphics[width=5.5cm]{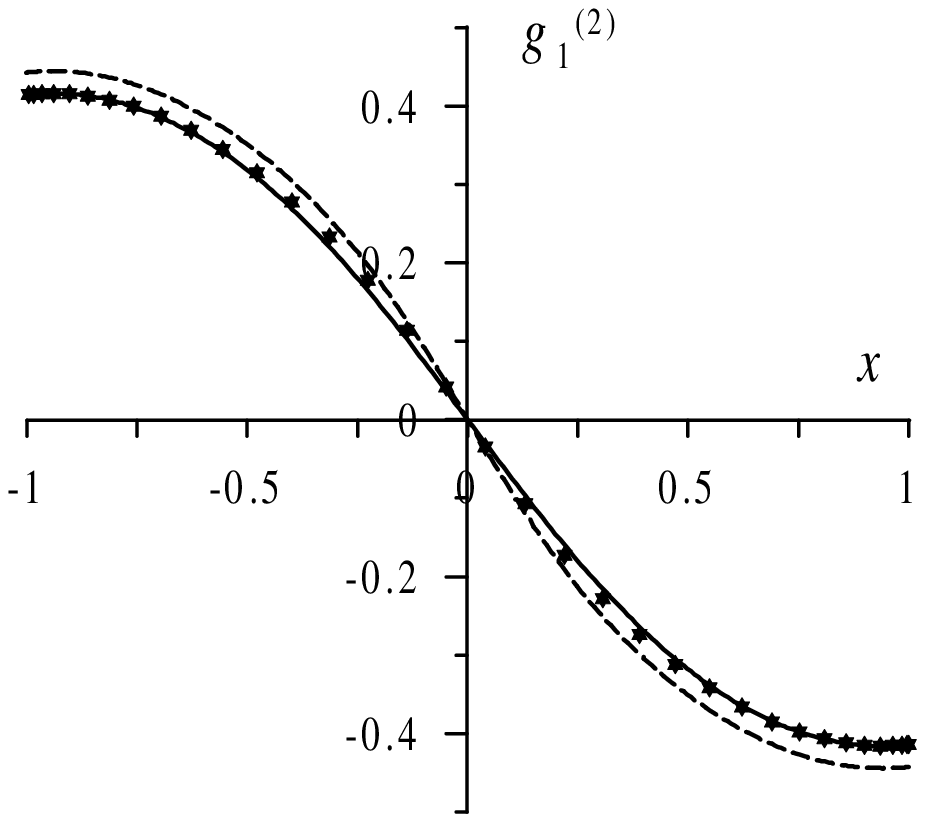}
\caption{Function $g_1^{(2)}$ found numerically (solid) and evaluated in the
``zero-derivative'' approximation (dashed). Asterisks show a corrected calculation with
allowance for the coefficient $D_2$, see Eq.~(\protect\ref{eq:49}). \label{fig:04}}
\end{figure}

It is instructive to compare the approximate expressions~(\ref{eq:45}) with the numerical
results obtained without simplification of $g_1^{(0)}$. To be specific, we consider the case
when probing field is applied along the particle easy axis $\bm{n}$. Then Eqs.~(\ref{eq:43})
become one-dimensional and the second of them writes
\begin{equation}                                                        \label{eq:46}
\frac{dg_{1}^{(n)}}{dx}=g_{1}^{(n-1)}.
\end{equation}
Its ``zero-derivative'' solutions up to the second order follow from the first two lines of
Eqs.~(\ref{eq:45}):
\begin{equation}                                                        \label{eq:47}
g_{1}^{(1)}= \varphi_1\,x-\langle\,\varphi_1 x\rangle_0, \qquad g_{1}^{(2)}=
\textstyle{\frac{1}{2}}\,\varphi_1\, x^2-x \langle\,\varphi_1 x\rangle_0\,.
\end{equation}
In Figs.~\ref{fig:03} and \ref{fig:04} these functions are compared to the numerical solutions
of Eq.~(\ref{eq:46}). For our calculation, the most important is the behavior of those
functions near $x=\pm1$ since these regions yield the main contribution when integrated with
the weight function $\psi_0$. As one can see from the figures, the ``zero-derivative''
solution $g_{1}^{(1)}$ agrees well with the exact one, whilst $g_{1}^{(2)}$ deviates
significantly. This discrepancy is due to the change of the barrier height that occurs in the
second order with respect to the probing field amplitude, and manifests itself in all the even
orders of the perturbation expansion. Correction of solution~(\ref{eq:47}) makes the second
step of our procedure. For that we integrate Eq.~(\ref{eq:46}) two times by parts
and substitute there the ``zero-derivative'' form of $g_{1}^{(1)}$ from Eq.~(\ref{eq:47}):
\begin{equation}                                                        \label{eq:48}
g_{1}^{(2)}=\frac{1}{2}\,x^2\varphi_1
-x\langle\,x\varphi_1\,\rangle+\frac{1}{2}\int\,x^2\frac{d\varphi_1}{dx}dx,
\end{equation}
Thus one finds that the corrected $g_1^{(2)}$ differs from this of Eq.~(\ref{eq:47}) by adding
a step-wise [alike that of Eq.~(\ref{eq:23})] term
\begin{equation}                                                        \label{eq:49}
g_{1}^{(2)}=\frac{1}{2}\,x^2\varphi_1 -x\langle\,x\varphi_1\,\rangle+D_2\varphi_1,
\end{equation}
with the amplitude
\begin{equation}                                                        \label{eq:50}
D_2=\frac{1}{2}\int_0^1 x^2\frac{d\varphi_1}{dx} dx.
\end{equation}
We remark that the results of evaluation of the integrals $I_{2k}=\int_0^1
x^{2k}\,(d\varphi_1/dx)\,dx$ can be arranged in the table
\begin{equation}                                                        \label{eq:51}
\begin{array}{|c|c|c|c|}
\>\>k\>\> & \;\;\;0  \;\; & 1 & 2 \\ \hline%
I_{2k} & \;\;\;1 \;\; & \>1-G^{-1}\> & \>1-(1+1/2\sigma)\,G^{-1}\>
\end{array}.
\end{equation}
so that Eq.~(\ref{eq:50}) gives
\begin{equation}                                                        \label{eq:52}
D_2=\frac{1}{2}\,I_{2k}=\frac{G-1}{2G} =
\frac{1}{4\sigma}+\frac{1}{4\sigma^2}+\frac{5}{8\sigma^3}+ \frac{37}{16\sigma^4}+ \ldots
\end{equation}
Function $g_{1}^{(2)}$ corrected in such a way is shown in Fig.~\ref{fig:04} by asterisks. It
is seen that the corrected dependence with a fairly good accuracy follows the the numerically
obtained curve.

In a similar way one can prove that the corrected function $g_{1}^{(4)}$ has the form
\begin{eqnarray}                                                        \label{eq:53}
g_{1}^{(4)}=\frac{1}{24}\,\varphi_1\,x^4
-\frac{1}{6}\left[\langle\,\varphi_1\,x\,\rangle_0\,x^3-3x\,
\langle\,x^2\,\rangle_0\right] \\
-\frac{1}{6}x\,\langle\,\varphi_1\,x^3\,\rangle_0\,+ D_2g_{1}^{(2)}+ D_4\varphi_1,
                                                                        \nonumber
\end{eqnarray}
where the corrected function $g_{1}^{(2)}$ given by Eq.~(\ref{eq:49}) is used and
\begin{eqnarray}                                                        \label{eq:54}
D_4=\frac{1}{24}\,I_{4}-D_2^2=
-\frac{10\sigma G^2-22\sigma G+G+12\sigma}{48\sigma G^2}\\
=-\frac{1}{32\sigma^2}-\frac{1}{16\sigma^3}-\frac{5}{32\sigma^4}- \frac{29}{64\sigma^5}+\ldots
                                                                        \nonumber
\end{eqnarray}
In the general case, when the direction of the probing field does not coincide with the
particle anisotropy axis, the corrected functions $g_{1}^{(n)}$ still can be written as
\begin{widetext}
\begin{eqnarray}                                                        \label{eq:55}
g_{1}^{(2)}&=&\textstyle{\frac{1}{2}}\varphi_1\,(\bm{e}\cdot\bm{h})^2-
(\bm{e}\cdot\bm{h})\,\langle\,\varphi_1(\bm{e}\cdot\bm{h})\, \rangle_0+D_2\varphi_1,
                                                                        \nonumber\\
g_{1}^{(3)}&=&\textstyle{\frac{1}{6}}\left[\varphi_1\,(\bm{e}\cdot
\bm{h})^3-\langle\,\varphi_1(\bm{e}\cdot\bm{h})^3\,\rangle_0
\right]-\textstyle{\frac{1}{2}}\langle\,\varphi_1(\bm{e}\cdot\bm{h})\,
\rangle_0\,\left[(\bm{e}\cdot\bm{h})^2- \langle\,(\bm{e}\cdot \bm{h})^2\,\rangle_0\right]+ D_2
g_{1}^{(1)},
                                                                        \nonumber\\
g_{1}^{(4)}&=&\textstyle{\frac{1}{24}}\varphi_1\,(\bm{e}\cdot\bm{h})^4
-\textstyle{\frac{1}{6}}\langle\,\varphi_1\,(\bm{e}\cdot
\bm{h})\,\rangle_0\,\left[(\bm{e}\cdot\bm{h})^3-3(\bm{e}
\cdot\bm{h})\,\langle\,(\bm{e}\cdot\bm{h})^2\,\rangle_0\right]
                                                                        \nonumber\\
& &\hspace*{8mm} -\textstyle{\frac{1}{6}}(\bm{e}\cdot\bm{h})\,\langle\,\varphi_1\,(\bm{e}\cdot
\bm{h})^3\,\rangle_0+D_2 g_{1}^{(2)}+D_4 \varphi_1\,.
\end{eqnarray}
\end{widetext}
But since Eqs.~(\ref{eq:43}) cannot be reduced to a form like Eq.~(\ref{eq:46}), the
correcting coefficients $D_2$ and $D_4$ cannot be presented in a closed form. In this case the
corrected solutions taking into account the behavior of function $\varphi_1$ around zero are
built up as power series near $x=0$; such a procedure for the coefficients $D_2$ and $D_4$ is
described in Appendix \ref{sec:AppB}.

\subsection{Dynamic probing field \label{sec:3.2}}
To obtain the dynamic susceptibilities, one has to find the distribution function $W$ in the
oscillating probing field $\xi\exp(i\omega t)$. For this situation the kinetic equation
(\ref{eq:02}) takes the form
\begin{equation}                                                        \label{eq:56}
\left(2\tau_D\frac{\partial}{\partial t}+\widehat{L}\right)\,W(t) =\xi\widehat{V}\,e^{i\omega
t}\,W(t),
\end{equation}
where the operators $\widehat{L}$ and $\widehat{V}$ have been introduced in above. Assuming
that the exciting field amplitude is not too high, we expand the steady-state oscillatory
solution of Eq.~(\ref{eq:56}) in a power series with respect to $\xi$:
\begin{equation}                                                        \label{eq:57}
W(t)=\sum_{n=0}\,\xi^nW^{(n)}\,e^{in\omega t}.
\end{equation}
Note that, mathematically, representation~(\ref{eq:57}) is not complete. Indeed, in a general
case the exact amplitude of the $n\omega$-mode must contain, along with the contribution
$\sim\xi^n$, an infinite set of terms $\sim\xi^{n+2}$, $\xi^{n+4}$, etc. However, in a weak
field limit $\xi<1$ the terms with higher powers are of minor importance so that the main
contribution to the magnetization response signal filtered at the frequency $n\omega$ is
proportional to $\xi^n$.

Substituting Eq.~(\ref{eq:57}) in (\ref{eq:56}) we arrive at the recurrence set
\begin{equation}                                                        \label{eq:58}
\left(2in\omega\tau_D+\widehat{L}\right)\,W^{(n)} =\widehat{V}\,W^{(n-1)},
\end{equation}
that we solve sequentially starting from $n=1$. At the first step the function in the
right-hand side corresponds to the equilibrium case ($\xi=0$). Therefore, $W^{(0)}=\psi_0$,
where the latter function is defined by Eq.~(\ref{eq:11}) and is frequency-independent.
Combining Eq.~(\ref{eq:42}) written down for $\beta=0$ and $n=1$ and Eq.~(\ref{eq:58}), we
eliminate the operator $\widehat{V}$ and get
\begin{equation}                                                        \label{eq:59}
\left(2i\omega\tau_D+\widehat{L}\right)\,W^{(1)}=\widehat{L} f_0^{(1)}.
\end{equation}
Now we expand the functions subjected to operator $\widehat{L}$ with respect to the set
$\{\psi_k\}$ of its eigenfunctions, see Eq.~(\ref{eq:06}):
\begin{equation}                                                        \label{eq:60}
W^{(1)}=\sum c^{(1)}_j(\omega)\psi_j\, \qquad
f_0^{(1)}=\sum\left(\varphi_j\bigl|\,f_0^{(1)}\right)\,\psi_j\,;
\end{equation}
here $(\varphi|f)$ denotes functional scalar multiplication, i.e., the integral of the product
$\varphi f$ over all the orientations of $\bm{e}$. Substitution of Eq.~(\ref{eq:60}) in
(\ref{eq:59}), multiplication of it from the left by $\varphi_k$ and integration, render the
expansion coefficient as
\begin{equation}                                                        \label{eq:61}
c_k^{(1)}(\omega)=\left(\varphi_k\bigl|f_0^{(1)}\right)\, \left[1+i\omega\tau_k\right]^{-1},
\end{equation}
where the reference relaxation times are defined by Eq.~(\ref{eq:26}).

In the low-frequency limit only $\omega\tau_1$ is set to be non-zero whilst all the higher
modes are taken at equilibrium ($\omega\tau_k=0$). Thence, when constructing $W^{(1)}$ via
Eq.~(\ref{eq:60}), by adding and subtracting a term with $c_1^{(1)}(0)$, one can present the
first-order solution in the form
\begin{equation}                                                        \label{eq:62}
W^{(1)}=f_0^{(1)}-\frac{i\omega\tau_1}{1+i\omega\tau_1}\, \left(\varphi_1\bigl|
f_0^{(1)}\right)\,\psi_1,
\end{equation}
where $f_0^{(1)}$, as seen from Eq.~(\ref{eq:59}), is the equilibrium solution for the same
value of the field amplitude $\xi$. We remind that the functions without upper index belong to
the fundamental set defined by Eqs.~(\ref{eq:06}) whereas those with an upper index are
evaluated in the framework of the perturbation scheme described in Sec.~\ref{sec:3.1}.

In the next order in $\xi$ the function $W^{(1)}$ is substituted in the right-hand side of
Eq.~(\ref{eq:58}) and through a procedure alike to that leading to
Eqs.~(\ref{eq:59})--(\ref{eq:61}), the function $W^{(2)}$ is found. We carry on this cycle up
to $k=5$. The results write
\begin{equation}                                                        \label{eq:63}
W^{(2)}=f^{(2)}_0-\frac{i\omega\tau_1}{1+i\omega\tau_1}
\left(\varphi_1\bigl|f^{(1)}_0\right)f_1^{(1)}\,,
\end{equation}
\begin{widetext}
\begin{eqnarray}                                                        \label{eq:64}
W^{(3)}=f^{(3)}_0+\left[\left(\varphi_1\bigl|f^{(1)}_0\right)
\left(\varphi_1\bigl|f_1^{(2)}\right)-\left(\varphi_1\bigl|f^{(3)}_0
\right)\right]f_1^{(0)}-\left(\varphi_1\bigl|f^{(1)}_0\right)f_1^{(2)}
                                                                        \nonumber \\
+\frac{1}{1+i\omega\tau_1}\left[\left(\varphi_1\bigl|f^{(1)}_0\right)\,
f_1^{(2)}-\textstyle{\frac{3}{2}}\left(\varphi_1\bigl|f_0^{(1)}\right)
\left(\varphi_1\bigl|f_1^{(2)}\right)f_1^{(0)}\right]
                                                                        \nonumber\\
+\frac{1}{1+3i\omega\tau_1}\left[\left(\varphi_1\bigl|f^{(3)}_0\right)
+\textstyle{\frac{1}{2}}\left(\varphi_1\bigl|f^{(1)}_0\right)\left(\varphi_1\bigl|\,
f_1^{(2)}\right)\right]f_1^{(0)}\,,
\end{eqnarray}

\begin{eqnarray}                                                        \label{eq:65}
W^{(4)}=f_0^{(4)}+\left[\left(\varphi_1\bigl|f^{(1)}_0\right)
\left(\varphi_1\bigl|f_1^{(2)}\right)-\left(\varphi_1\bigl|f^{(3)}_0 \right)\right]
f_1^{(1)}-\left(\varphi_1\bigl|f_0^{(1)}\right)f_1^{(3)}
                                                                        \nonumber \\
+\frac{1}{1+i\omega\tau_1}\left[\left(\varphi_1\bigl|f^{(1)}_0\right)\,
f_1^{(3)}-\textstyle{\frac{3}{2}}\left(\varphi_1\bigl|f^{(1)}_0\right)
\left(\varphi_1\bigl|f_1^{(2)}\right)f_1^{(1)}\right]
                                                                        \nonumber \\
+\frac{1}{1+3i\omega\tau_1}\left[\left(\varphi_1\bigl|f^{(3)}_0\right)
+\textstyle{\frac{1}{2}}\left(\varphi_1\bigl|f^{(1)}_0\right)\left(\varphi_1\bigl|\,
f_1^{(2)}\right)\right]f_1^{(1)}\,,
\end{eqnarray}

\begin{eqnarray}                                                        \label{eq:66}
W^{(5)}&=&f^{(5)}_0-\left(\varphi_1\bigl|f^{(5)}_0\right)f_1^{(0)}
+\left(\varphi_1\bigl|f^{(1)}_0\right)\left[\left(\varphi_1\bigl|\,
f_1^{(4)}\right)f_1^{(0)}-f_1^{(4)}\right] \hspace*{45mm}
                                                                        \nonumber \\
&+&\left[\left(\varphi_1\bigl|f^{(1)}_0\right)\left(\varphi_1\bigl|\,
f_1^{(2)}\right)-\left(\varphi_1\bigl|f^{(3)}_0\right)\right]\left[
f_1^{(2)}-\left(\varphi_1\bigl|f_1^{(2)}\right)f_1^{(0)}\right]
                                                                        \nonumber \\
&+&\frac{1}{1+i\omega\tau_1}\left(\varphi_1\bigl|f^{(1)}_0\right)\left\{
f_1^{(4)}\!\!-\!\textstyle{\frac{3}{2}}\left(\varphi_1\bigl|f_1^{(2)}\right)f_1^{(2)}
+\left[\textstyle{\frac{15}{8}}\left(\varphi_1\bigl|f_1^{(2)}\right)^2\!\!-
\!\!\textstyle{\frac{5}{4}}\left(\varphi_1\bigl|f_1^{(4)}\right)\right]f_1^{(0)}\right\}
                                                                        \nonumber \\
&+&\frac{1}{1+3i\omega\tau_1}\left[\left(\varphi_1\bigl|f^{(3)}_0\right)
+\textstyle{\frac{1}{2}}\left(\varphi_1\bigl|f^{(1)}_0\right)\left(\varphi_1\bigl|\,
f_1^{(2)}\right)\right]\left[f_1^{(2)}-\textstyle{\frac{5}{2}}\left(\varphi_1\bigl|\,
f_1^{(2)}\right)f_1^{(0)}\right]
                                                                        \nonumber \\
&+&\frac{1}{1+5i\omega\tau_1}\left[\left(\varphi_1\bigl|f^{(5)}_0\right)+
\textstyle{\frac{1}{4}}\left(\varphi_1\bigl|f^{(1)}_0\right)\left(\varphi_1\bigl|\,
f_1^{(4)}\right)+\textstyle{\frac{3}{8}}\left(\varphi_1\bigl|f^{(1)}_0\right)
\left(\varphi_1\bigl|f_1^{(2)}\right)^2\right.
                                                                        \nonumber \\
&&\hspace*{78mm}\left.+\textstyle{\frac{3}{2}}\left(\varphi_1\bigl|f^{(3)}_0\right)
\left(\varphi_1\bigl|f_1^{(2)}\right)\right]f_1^{(0)}.
\end{eqnarray}
\end{widetext}
We remark an important feature of Eqs.~(\ref{eq:63})--(\ref{eq:66}): they do not contain
dispersion factors of even orders. This ensures that the frequency dependence of the full
distribution function $W$ incorporates only dispersion factors with odd multiples of the basic
frequency. Qualitatively, this is the result of absence of the interwell mode for the
statistical moments of even orders. Technically, it is due to vanishing of the products
$\left(\varphi_1\bigl|f_k^{(\ell)}\right)$ entering Eqs.~(\ref{eq:62})--(\ref{eq:66}) if the
sum $k+\ell$ is even. This rule follows immediately from combination of the oddity of
$\varphi_1$, see Sec.~\ref{sec:2}, with the parity properties of the functions $f_k^{(\ell)}$
introduced in Sec.~\ref{sec:3.1}.

For actual calculations one needs the values of the scalar products entering
Eqs.~(\ref{eq:62})--(\ref{eq:66}). In Appendix~\ref{sec:AppC} we obtain their representations
in terms of the moments $Q_k$ and $S_k$ of the functions $\psi_0$ and $\psi_1$, respectively.
The procedures of asymptotic expansion of $Q_k$ and $S_k$ are given in
Appendix~\ref{sec:AppA}.

\section{Dynamic susceptibilities \label{sec:4}}
The set of magnetic susceptibilities of an assembly of non-interacting particles with the
number density $c$ is defined by the relation
\begin{equation}                                                        \label{eq:67}
M=\chi^{(1)}H+\chi^{(3)}H^3+\chi^{(5)}H^5+\ldots
\end{equation}
that describes the magnetization of the system in the direction of the probing field
$\bm{H}=H\bm{h}$. Therefore, of all the components of the corresponding susceptibility
tensors, we retain the combinations that determine the response in the direction of the
probing field. With representation~(\ref{eq:57}) for the distribution function, this
magnetization component takes the form
\begin{eqnarray}                                                        \label{eq:68}
&& M=cIv\langle\,(\bm{e}\cdot\bm{h})\,\rangle                              \\
&& =c\sum_{n=1}\,H^n\,\frac{I^ {n+1}v^{n+1}}{(kT)^n}e^{in\omega t}\int(\bm{e}
\cdot\bm{h})\,W^{(n)}\,d \bm{e},                                        \nonumber
\end{eqnarray}
and the susceptibilities can be found by a direct comparison with Eq.~(\ref{eq:67}). In other
words, the set of $\chi^{(n)}$ is expressed through the perturbation functions $W^{(n)}$ found
in the preceding section. Therefore, evaluation of $\chi^{(n)}$ becomes, although tedious, but
simple procedure. Remarkably, the final expressions come out rather compact.
\subsection{Linear susceptibility}
The resulting expression can be presented in the form
\begin{equation}                                                        \label{eq:69}
\chi^{(1)}_{\omega}=\chi_0^{(1)}\left(B_0^{(1)}+\frac{B_1^{(1)}}{1+i\omega \tau_1}\right),
\qquad \chi_0^{(1)}=\frac{cI^2v^2}{3kT},
\end{equation}
which follows from substituting Eq.~(\ref{eq:62}) in (\ref{eq:68}). Each of the two
frequency-independent coefficients $B^{(1)}$, being the result of statistical averaging over
the orientational variable $\bm{e}$, see Appendix~\ref{sec:AppC}, expands into a series of
Legendre polynomials with respect to $\beta$, the angle between the direction $\bm{h}$ of the
probing field and the particle easy axis $\bm{n}$. This can be written as
\begin{eqnarray}                                                        \label{eq:70}
&& \begin{cases}
B_0^{(1)}=b_{00}^{(1)}+b_{02}^{(1)}P_2(\cos\beta),& \\[5pt]
B_1^{(1)}=b_{10}^{(1)}+b_{12}^{(1)}P_2(\cos\beta),&
\end{cases}
                                                                        \\[6pt]
&& \left(
\begin{array}{cc}
b_{00}^{(1)} & b_{02}^{(1)} \\
b_{10}^{(1)} & b_{12}^{(1)}
\end{array}
\right)=
\left(
\begin{array}{cc}
1-Q_1^2 & \hspace*{5mm} 2S_2-2Q_1^2 \\
Q_1^2 & \hspace*{5mm} 2Q_1^2                                            \nonumber
\end{array}
\right).
\end{eqnarray}
Definitions of functions $Q_1$ and $S_2$ and their explicit asymptotic representations are
given in Appendix~\ref{sec:AppA}. The asymptotic series for the coefficients
$b_{\alpha\beta}^{(1)}$ derived on the base of expansion~(\ref{eq:12}) and Eq.~(\ref{eq:37})
are
\begin{widetext}
\begin{eqnarray}                                                        \label{eq:71}
&& b_{02}^{(1)}=-\frac{1}{\sigma}+\frac{1}{4\sigma^3}+\frac{13}{8\sigma^4}+
\frac{165}{16\sigma^5}
+\frac{2273}{32\sigma^6}+\frac{34577}{64\sigma^7}+\frac{581133}{128\sigma^8}+\ldots
                                                                        \\[7pt]
&& b_{10}^{(1)}=1-\frac{1}{\sigma}-\frac{3}{4\sigma^2}-\frac{2}{\sigma^3}-
\frac{31}{4\sigma^4} -\frac{153}{4\sigma^5}-\frac{3629}{16\sigma^6}-
\frac{1564}{\sigma^7}-\frac{785931}{64\sigma^8}+\ldots \nonumber
\end{eqnarray}
\end{widetext}
The other components, namely, $b_{00}^{(1)}$ and $b_{12}^{(1)}$, may be constructed
straightforwardly using their relations with the given ones, see Eqs.~(\ref{eq:70}). For a
random system, that is for an assembly of non-interacting particles with a chaotic
distribution of the anisotropy axes, the average of any Legendre polynomial is zero, so that
$\overline{B^{(1)}}_k=b^{(1)}_{k 0}$, and the linear dynamic susceptibility reduces to
\begin{equation}                                                        \label{eq:72}
\overline{\chi^{(1)}}_{\omega}=\chi_0^{(1)}\,\frac{1+i\omega\tau_1b_{00}}{1+ i\omega\tau_1},
\end{equation}
that is the asymptotic representation of the full expression given by formula~(39) of
Ref.~\cite{RaSt-PRB:97}.

\subsection{Cubic susceptibility}
As follows from definitions~(\ref{eq:67}) and (\ref{eq:68}), the third-order susceptibility is
defined through the response at the triple frequency that at weak $H$ scales as $H^3$.
Performing calculations along the same scheme as for $\chi^{(1)}$, one arrives at the
sum-of-relaxators representation
\begin{eqnarray}                                                        \label{eq:73}
&& \chi_{3\omega}^{(3)}=\frac{1}{4}\,\chi_0^{(3)}\left(B_0^{(3)}+\frac{B_1^{(3)}}
{1+i\omega\tau_1}+\frac{B_3^{(3)}}{1+3i\omega\tau_1}\right),            \\
&& \chi_0^{(3)}=\frac{cI^4v^4}{(kT)^3},                                    \nonumber
\end{eqnarray}
where the coefficients expand as
\begin{equation}                                                        \label{eq:74}
B_k^{(3)}=b_{k0}^{(3)}+b_{k2}^{(3)}P_2(\cos\beta)+ b_{k4}^{(3)}P_4(\cos\beta), \quad k=0,1,3
\ldots
\end{equation}
up to the fourth Legendre polynomial in $\cos\beta$.

The explicit expansions for the amplitudes $b^{(3)}_{\alpha\beta}$ are
\begin{widetext}
{\scriptsize
\begin{eqnarray}                                                        \label{eq:75}
b_{00}^{(3)}&=&\frac{1}{30\sigma^3}+\frac{47}{240\sigma^4}
+\frac{49}{40\sigma^5}+\frac{815}{96\sigma^6}+\frac{7837}{120\sigma^7}
+\frac{355391}{640\sigma^8}+\ldots
                                                                        \\[4pt]
b_{02}^{(3)}&=&\frac{1}{42\sigma^3}+\frac{2}{21\sigma^4}
+\frac{4}{7\sigma^5}+\frac{1385}{336\sigma^6}
+\frac{11231}{336\sigma^7}+\frac{19083}{64\sigma^8}+\ldots
                                                                        \nonumber \\[4pt]
b_{04}^{(3)}&=&-\frac{2}{35\sigma^3}-\frac{8}{35\sigma^4}
-\frac{41}{35\sigma^5}-\frac{50}{7\sigma^6}
-\frac{1756}{35\sigma^7}-\frac{63749}{160\sigma^8}+\ldots
                                                                        \nonumber \\[4pt]
b_{10}^{(3)}&=&\frac{1}{15}-\frac{1}{6\sigma}-\frac{23}{240\sigma^2}
-\frac{61}{192\sigma^3}-\frac{1357}{960\sigma^4}
-\frac{235447}{30720\sigma^5}-\frac{11962691}{245760\sigma^6}
-\frac{694849241}{1966080\sigma^7}-\frac{15133953221}{5242880\sigma^8}
+\ldots
                                                                        \nonumber \\[4pt]
b_{12}^{(3)}&=&\frac{13}{84}-\frac{65}{168\sigma}-\frac{25}{168\sigma^2}
-\frac{863}{1344\sigma^3}-\frac{3931}{1344\sigma^4}
-\frac{698911}{43008\sigma^5}-\frac{35309123}{344064\sigma^6}
-\frac{2061480665}{2752512\sigma^7}-\frac{45071465669}{7340032\sigma^8}
+\ldots
                                                                        \nonumber \\[4pt]
b_{14}^{(3)}&=&\frac{1}{35}-\frac{1}{14\sigma}+\frac{2}{35\sigma^2}
-\frac{1}{112\sigma^3}-\frac{73}{560\sigma^4}
-\frac{17033}{17920\sigma^5}-\frac{1007549}{143360\sigma^6}
-\frac{64390439}{1146880\sigma^7}-\frac{4493994417}{9175040\sigma^8}
+\ldots
                                                                        \nonumber \\[4pt]
b_{30}^{(3)}&=&-\frac{2}{15}+\frac{3}{10\sigma}
+\frac{1}{16\sigma^2}+\frac{337}{960\sigma^3}
+\frac{499}{320\sigma^4}+\frac{85309}{10240\sigma^5}
+\frac{2563751}{49152\sigma^6}+\frac{245269747}{655360\sigma^7}
+\frac{47628510799}{15728640\sigma^8}+\ldots
                                                                        \nonumber \\[4pt]
b_{32}^{(3)}&=&-\frac{29}{84}+\frac{43}{56\sigma}+\frac{11}{56\sigma^2}
+\frac{1279}{1344\sigma^3}+\frac{1881}{448\sigma^4}
+\frac{320765}{14336\sigma^5}+\frac{48133699}{34406\sigma^6}
+\frac{920146163}{91750\sigma^7}+\frac{178560431695}{22020096\sigma^8}
+\ldots
                                                                        \nonumber \\[4pt]
b_{34}^{(3)}&=&-\frac{11}{105}+\frac{47}{210\sigma}+\frac{2}{21\sigma^2}
+\frac{559}{1680\sigma^3}+\frac{2419}{1680\sigma^4}
+\frac{409499}{53760\sigma^5}+\frac{4080395}{86016\sigma^6}
+\frac{1166954357}{3440640\sigma^7}+\frac{75334335763}{27525120\sigma^8} +\ldots
                                                                        \nonumber
\end{eqnarray}}
For a random system, the averages of Legendre polynomials drop out and $\overline{B_k^{(3)}}=
b^{(3)}_{k0}$. With respect to formalism constructed in Ref.~\cite{RaSt-PRB:97}, the above
expressions yield the asymptotic representations for formulas~(42) and (43) there.

\begin{figure}
\includegraphics[width=11cm]{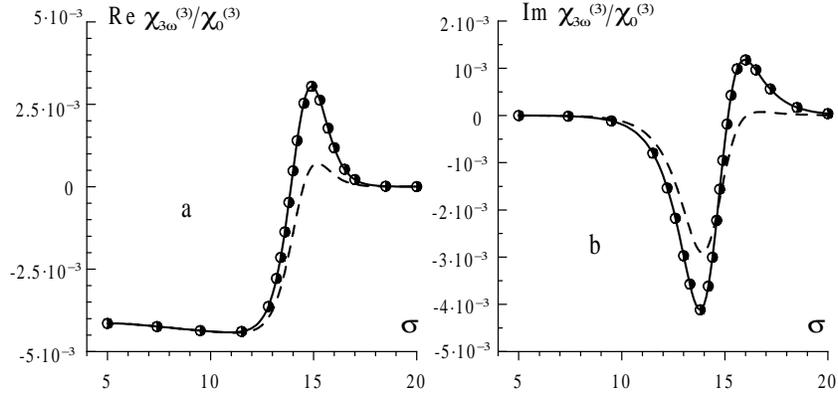}
\caption{Real (a) and imaginary (b) components of the cubic susceptibility of a
superparamagnetic assembly with coherently aligned easy axes; the direction of the probing
field is tilted with respect to the alignment axis at $\cos\beta=0.5$; the dimensionless
frequency is $\omega\tau_0=10^{-6}$. Solid lines show the proposed asymptotic formulas taken
with the accuracy $\sigma^{-3}$, circles present the result of numerically-exact evaluation,
dashed lines correspond to the ``zero-derivative'' approximation (\protect\ref{eq:45}). The
discrepancy of the curves is commented in the text after Eq.~(\ref{eq:B12}). \label{fig:05}}
\end{figure}

\subsection{Fifth-order susceptibility}
The susceptibility of the fifth order writes in an expectable way as a sum of three
relaxators:
\begin{equation}                                                        \label{eq:76}
\chi_{5\omega}^{(5)}=\frac{1}{16}\chi_0^{(5)} \left(B_0^{(5)}+\frac{B_1^{(5)}}
{1+i\omega\tau}+\frac{B_3^{(5)}}{1+3i\omega\tau}+\frac{B_5^{(5)}}{1+5i\omega\tau}\right),
\qquad \chi_0^{(5)}= \frac{cI^6v^6}{(kT)^5},
\end{equation}
with the coefficients
\begin{equation}                                                        \label{eq:77}
B_k^{(5)}=b_{k0}^{(5)}+b_{k2}^{(5)}P_2(\cos\beta)
+b_{k4}^{(5)}P_4(\cos\beta)+b_{k6}^{(5)}P_6(\cos\beta) \qquad k=0,1,3,5.
\end{equation}
The explicit asymptotic series are
{\scriptsize
\begin{eqnarray}                                                        \label{eq:78}
b_{00}^{(5)}&=&\frac{1}{80\sigma^5}+\frac{367}{2240\sigma^6}
+\frac{123}{70\sigma^7}+\frac{41233}{2240\sigma^8}+\ldots
                                                                 \nonumber \\
b_{10}^{(5)}&=&\frac {1}{96}-\frac {19}{420\sigma}+\frac{1}{120\sigma^2}
-\frac{65}{1792\sigma^3}-\frac{79}{336\sigma^4} -\frac{85913}{57344\sigma^5}-\frac
{72636131}{6881280\sigma^6}
-\frac{4543038053}{55050240\sigma^7}-\frac{14938598691}{20971520\sigma^8} +\ldots
                                                                  \nonumber \\
b_{30}^{(5)}&=&-\frac{47}{560}+\frac{11}{35\sigma}-\frac {29}{280\sigma^2} +\frac
{437}{1920\sigma^3}+\frac{5473}{4480\sigma^4} +\frac
{1046209}{143360\sigma^5}+\frac{169435283}{3440640\sigma^6}
+\frac{684614895}{1835008\sigma^7}+\frac{230861266333}{73400320\sigma^8} +\ldots
                                                                 \nonumber \\
b_{50}^{(5)}&=&\frac{311}{3360}-\frac{137}{420\sigma}+\frac{13}{105\sigma^2}
-\frac{5911}{26880\sigma^3}-\frac{2141}{1920\sigma^4}
-\frac{1874309}{286720\sigma^5}-\frac{299470403}{6881280\sigma^6}
-\frac{17964831133}{55050240\sigma^7}-\frac{400677748549}{146800640\sigma^8} +\ldots
                                                                 \nonumber \\
b_{02}^{(5)}&=&\frac{1}{112\sigma^5}+\frac{3}{28\sigma^6}
+\frac{507}{448\sigma^7}+\frac{5377}{448\sigma^8} +\ldots
                                                                 \nonumber \\
b_{12}^{(5)}&=&\frac{13}{504}-\frac {19}{168\sigma}+\frac {23}{672\sigma^2}
-\frac{737}{8064\sigma^3}-\frac{2959}{5376\sigma^4}
-\frac{99733}{28672\sigma^5}-\frac{50499149}{2064384\sigma^6}
-\frac{350973527}{1835008\sigma^7}-\frac{72765921299}{44040192\sigma^8} +\ldots
                                                                 \nonumber \\
b_{32}^{(5)}&=&-\frac {5}{21}+\frac{149}{168\sigma}-\frac{193}{672\sigma^2}
+\frac{5245}{8064\sigma^3}+\frac{18677}{5376\sigma^4}
+\frac{1785635}{86016\sigma^5}+\frac{289305193}{2064384\sigma^6}
+\frac{5846947361}{5505024\sigma^7}+\frac{394448762615}{44040192\sigma^8} +\ldots
                                                                 \nonumber \\
b_{52}^{(5)}&=&\frac{139}{504}-\frac{27}{28\sigma}+\frac{109}{336\sigma^2}
-\frac{1343}{2016\sigma^3}-\frac{9203}{2688\sigma^4}
-\frac{431321}{21504\sigma^5}-\frac{9839105}{73728\sigma^6}
-\frac{196654913}{196608\sigma^7}-\frac{30690812563}{3670016\sigma^8} +\ldots
                                                                 \nonumber \\
b_{04}^{(5)}&=&-\frac{3}{140\sigma^5}-\frac{1563}{6160\sigma^6}
-\frac{7767}{3080\sigma^7}-\frac{613353}{24640\sigma^8} +\ldots
                                                                 \nonumber \\
b_{14}^{(5)}&=&\frac{15}{2464}-\frac{183}{6160\sigma}+\frac{713}{24640\sigma^2}
-\frac{433}{19712\sigma^3}-\frac{409}{4928\sigma^4}
-\frac{319665}{630784\sigma^5}-\frac{8222083}{2293760\sigma^6}
-\frac{5744848239}{201850880\sigma^7} -\frac{403943151013}{1614807040\sigma^8}+\ldots
                                                                  \nonumber \\
b_{34}^{(5)}&=&-\frac{29}{280}+\frac{293}{770\sigma}-\frac{47}{385\sigma^2}
+\frac{7081}{24640\sigma^3}+\frac{74647}{49280\sigma^4}
+\frac{7137293}{788480\sigma^5}+\frac{385804437}{6307840\sigma^6}
+\frac{4682760003}{10092544\sigma^7} +\frac{1580817298041}{403701760\sigma^8}+\ldots
                                                                 \nonumber \\
b_{54}^{(5)}&=&\frac{1713}{12320}-\frac{2929}{6160\sigma}
+\frac{2551}{24640\sigma^2}-\frac{34863}{98560\sigma^3}
-\frac{92061}{49280\sigma^4}-\frac{34432191}{3153920\sigma^5}
-\frac{23756287}{327680\sigma^6}-\frac{15647080587}{28835840\sigma^7}
-\frac{7317549380671}{1614807040\sigma^8}+\ldots
                                                                 \nonumber \\
b_{06}^{(5)}&=&-\frac{1}{616\sigma^6}-\frac{3}{77\sigma^7} -\frac{1467}{2464\sigma^8}+\ldots
                                                                 \nonumber \\
b_{16}^{(5)}&=&-\frac{1}{1584}+\frac{1}{1848\sigma}+\frac {7}{1056\sigma^2}
+\frac{337}{88704\sigma^3}+\frac{53}{1848\sigma^4}
+\frac{51433}{315392\sigma^5}+\frac{2188103}{2064384\sigma^6}
+\frac{471762913}{60555264\sigma^7} +\frac{4428495037}{69206016\sigma^8}+\ldots
                                                                 \nonumber \\
b_{36}^{(5)}&=&-\frac{1}{84}+\frac{10}{231\sigma}-\frac {17}{924\sigma^2}
+\frac{103}{3168\sigma^3}+\frac{2489}{14784\sigma^4}
+\frac{236615}{236544\sigma^5}+\frac{38344237}{5677056\sigma^6}
+\frac{776232845}{15138816\sigma^7} +\frac{52467158027}{121110528\sigma^8}+\ldots
                                                                 \nonumber \\
b_{56}^{(5)}&=&\frac{1207}{55440}-\frac{661}{9240\sigma}
+\frac{17}{12320\sigma^2}-\frac{5525}{88704\sigma^3}
-\frac{1169}{3520\sigma^4}-\frac{9116467}{4730880\sigma^5}
-\frac{131486063}{10321920\sigma^6}-\frac{1918435847}{20185088\sigma^7}
-\frac{639291980689}{807403520\sigma^8}+\ldots,
                                                                 \nonumber \\
\quad
\end{eqnarray}}
\end{widetext}
and for a random system, as for the lower orders, $\overline{B_k^{(5)}}=b^{(0)}_{k0}$.

\begin{figure}
\includegraphics[width=5.5cm]{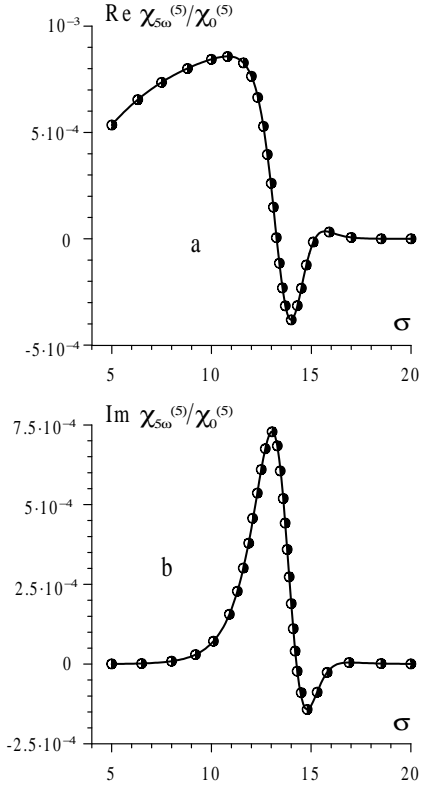}
\caption{Real (a) and imaginary (b) components of the fifth-order susceptibility of a random
superparamagnetic assembly; the dimensionless frequency is $\omega\tau_0=10^{-6}$. Solid lines
show the proposed asymptotic formulas with the accuracy $\sigma^{-3}$, circles present the
result of a numerical evaluation \label{fig:06}}
\end{figure}

\section{Discussion}%
\label{sec:5}%
The above derived formulas despite their hefty look are very practical. Indeed, they present
the nonlinear initial susceptibilities of a superparamagnetic particulate medium as analytical
expressions of arbitrary accuracy. With respect to the frequency dependence they give the
exact full structure of the susceptibility and prove that it is very simple thus putting
former intuitive considerations on a solid ground. This makes our formulas a handy tool for
asymptotic analysis. Yet more convenient they are for numerical work because with their use
the difficult and time-consuming procedure of solving the differential equations is replaced
by a plain summation of certain power series. For example, if to employ
Eqs.~(\ref{eq:72})--(\ref{eq:78}), a computer code that fits simultaneously experimental data
on linear and a set of nonlinear susceptibilities taking into account the particle
polydispersity of any kind (easy axes directions, activation volume, anisotropy constants)
becomes a very fast procedure.

Graphic examples justifying our claims are presented in Figs.~\ref{fig:05} and \ref{fig:06},
where the components of two nonlinear complex susceptibilities are plotted as the functions of
the parameter $\sigma$. For a given sample, $\sigma$ in a natural way serves as a
dimensionless inverse temperature. In those figures, the solid lines correspond to the
above-proposed asymptotic formulas where we keep the terms up to $\sigma^{-3}$. The circles
show the results of numerically-exact solutions obtained by the method described in
Ref.~\onlinecite{RaSt-PRB:97}. Note that even at $\sigma\sim5$ the accuracy is still rather
high.

The model that may be called the predecessor of the afore-derived results was proposed in
Ref.~\onlinecite{KlYa-JMMM:98}. There, the authors calculated the initial susceptibilities up
to the seventh order having replaced a superparamagnetic assembly by a two-level macrospin
system. The interrelation between the present work and Ref.~\onlinecite{KlYa-JMMM:98} closely
resembles the situation with the evaluation of the rate of a superparamagnetic process. First
in 1949 N\'eel\cite{Neel-AGCR:49} and then, ten years later, Brown\cite{Brown-JAP:59} had
evaluated the superparamagnetic time in the framework of a two-level model. In such a
framework, one allows for the magnetic moment flips but totally neglects its possible
diffusion over energetically less-favorable directions. In 1963 Brown\cite{Brown-PR:63} had
succeeded to overcome this artificial assumption and took into account the possibility for the
magnetic moment to wander over all $4\pi$ radians.

In the present case, the obtained $v/T$ dependencies of the nonlinear susceptibilities and
those from Ref.~\onlinecite{KlYa-JMMM:98} are qualitatively the same. Their most typical
feature is the double-peak shape. Quantitatively, however, the corresponding lines differ and
do not reduce to one another in any case. Indeed, as long as the temperature is finite
(whatever low), the configurational space for the unit vector $\bm{e}$ of the magnetic moment
is the full ($4\pi$-radian) solid angle; its reduction to just two directions along a
bidirectional axis could not be done otherwise than ``by hand''. This is exactly what the
two-level Ising-like model does: it forcibly imparts a quantum property (discrete spin
projections) to a macrospin assembly. From the calculational viewpoint, another essential
demerit of the results\cite{KlYa-JMMM:98} is that the coefficients in the susceptibility
formulas are not given in an analytic form. The authors propose to evaluate them by solving an
infinite set of recurrence equations. Hence, the procedure\cite{KlYa-JMMM:98} does not provide
any gain with respect to former ones neither in analytical considerations nor in constructing
fitting codes.

In the presented framework the results by Klik and Yao (including the analytical formulas for
them missing in Ref.~\onlinecite{KlYa-JMMM:98}) can be obtained immediately if to take the
function $\varphi_1$ in a step-wise form~(\ref{eq:23}) and not to allow for the corrections
caused by the finiteness of its derivative at $x=0$. In our terms this means to stop at
set~(\ref{eq:45}), i.e., ``zero-derivative'' solution, and not to go further. The emerging
error is however, uncontrollable and not at all small. As an illustration, in
Fig.~\ref{fig:05} we show the result obtained with this model (dashed lines) for the cubic
susceptibility $\chi_{3\omega}^{(3)}$ in a textured system where the particle common axis
$\bm{n}$ is tilted under the angle $\beta=\pi/3$ to the probing field. One can see that
deviations are substantial.

In Ref.~\onlinecite{RaSt-PRB:97} we have proposed, although without rigorous justification, a
formula for the cubic susceptibility of a random assembly
\begin{equation}                                                        \label{eq:79}
\chi_{3\omega}^{(3)}=-\frac{1}{4}\,\chi_0^{(3)}
\frac{(1+2S_2^2)(1-i\omega\tau_1)}{45(1+i\omega\tau_1)(1+3i\omega\tau_1)},
\end{equation}
that proved to be well adjusted for approximating the results of numerical calculations in all
the temperature interval and also appeared to be good for fitting experimental
data.\cite{SpFi-JMMM:01} Now we see that this very expression follows from
Eqs.~(\ref{eq:73})--(\ref{eq:75}) if to expand the coefficients $b_{i0}^{(3)}$ up to the
zeroth order with respect to $\sigma^{-1}$. This justifies Eq.~(\ref{eq:79}) as a formula
yielding a correct frequency dispersion of the cubic susceptibility of a random assembly at
low temperatures. The cause of its applicability at high temperatures is the exponential
dependence of $\tau_1$ on $\sigma$. Indeed, in the frequencies range $\omega\tau_0\ll1$, where
we work, the condition $\sigma\lesssim1$ means $\tau_1\rightarrow\tau_0$, and all the
dispersion factors in Eq.~(\ref{eq:79}) drop out. This transforms expression~(\ref{eq:79}) in
a correct static susceptibility that is also a true result. To avoid any confusion we remark
that Eq.~(\ref{eq:79}) differs from formula for $\chi_{3\omega}^{(3)}$ given in
Ref.~\cite{RaSt-PRB:97} by the coefficient ($-1/45$) due to the difference in definitions: in
Ref.~\cite{RaSt-PRB:97} it was included in $\chi_0^{(3)}$.

Applying the similar procedure to Eqs.~(\ref{eq:76})--(\ref{eq:78}) we get the expression for
the fifth-order susceptibility
\begin{widetext}
\begin{equation}                                                        \label{eq:80}
\chi_{5\omega}^{(5)}=\frac{1}{16}\chi_0^{(5)}
\frac{(2+12S_2^2+4S_2^3)}{945}\frac{1-\frac{21}{8}i\omega\tau_1- \frac{3}{4}\omega^2\tau_1^2}
{(1+i\omega\tau_1)(1+3i\omega\tau_1)(1+5i\omega\tau_1)},
\end{equation}
\end{widetext}
that, following the example of the already tested Eq.~(\ref{eq:79}), has high chances to be
a good approximation for $\chi_{5\omega}^{(5)}$ in the whole temperature interval. As we
have already ascertained in Ref.~\cite{RaSt-PRB:97}, the best interpolation expression for
the relaxation time in the susceptibility formulas is
\begin{displaymath}
\tau_1=\tau_D\frac{e^\sigma-1}{2\sigma}\>\left[\frac{\sigma}{1+\sigma}\sqrt{\frac{\sigma}{\pi}}
+2^{-\sigma-1}\right]^{-1},
\end{displaymath}
proposed in Refs.~\cite{CoCr-JMMM:94,CrCr-JAP:94}.

\section{Conclusions}
A consistent procedure yielding the integral relaxation time and initial nonlinear
susceptibilities for an assembly of non-interacting superparamagnetic particles is constructed
in the low-to-moderate temperature range. Starting from the micromagnetic kinetic equation
that describes intrinsic rotary diffusion of the particle magnetic moment, we obtain the
results in an analytical form. They are presented as asymptotic series with respect to the
dimensionless parameter $\sigma$ that is the uniaxial anisotropy barrier height scaled with
temperature. High-order expansion terms are easily accessible that allows to achieve any
desirable extent of accuracy. This is proven by comparison of the proposed approximation with
the numerically-exact results. The susceptibilities contain angular dependencies that allow
one to consider the particle assemblies with any extent of orientational texture---from
perfectly aligned to random. The new formulas stand closer to reality than those for a
two-level system and are to facilitate considerably both analytical and numerical calculations
in the theory of superparamagnetic relaxation in single-domain particles.

\begin{acknowledgments}
Partial financial support from the International Association for the Promotion of
Cooperation with Scientists from the New Independent States of the Former Soviet Union
(INTAS) under Grant No.\ 01--2341 and by Award No.\ PE--009--0 of the U.S. Civilian
Research \&{} Development Foundation for the Independent States of the Former Soviet Union
(CRDF) is gratefully acknowledged.

The package Maple V for PC used in the calculational work was obtained by Institute of
Continuous Media Mechanics in the framework of the EuroMath Network and Services for the New
Independent States - Phase II (EmNet/NIS/II) project funded by INTAS under grant No.\ IA-003.
\end{acknowledgments}

\appendix
\section{Evaluation of the expansion coefficients for eigenfunctions $\psi_0$ and $\psi_1$
         \label{sec:AppA}}
Both functions $\psi_0$ and $\psi_1$ are uniaxially symmetrical about the anisotropy axis
$\bm{n}$ and can be expanded in the Legendre polynomial series, see Eq.~(\ref{eq:32}):
\begin{eqnarray}                                                        \label{eq:A01}
& \psi_0=\textstyle{\frac{1}{2}}\sum\limits_{k=0}(2k+1)S_kP_k(x),
 \quad k=0,2,4,\ldots, \\
& \psi_1=\textstyle{\frac{1}{2}}\sum\limits_{k=1}(2k+1)Q_kP_k(x), \quad k=1,3,5,\ldots,
\nonumber
\end{eqnarray}
where in accordance with the parity properties of the eigenfunctions non-zero terms are
\begin{eqnarray}                                                        \label{eq:A02}
S_0=1, && S_k=\left(P_k(x)\bigl|\psi_0\right), \quad k=2,4,\ldots,\\[6pt]
&& Q_k=\left(P_k(x)\bigl|\,\psi_1\right), \quad k=1,3,5,\ldots .        \nonumber
\end{eqnarray}
Taking into account that $\psi_1=\psi_0\varphi_1$, where $\psi_0$ in a finite form is given by
Eq.~(\ref{eq:11}), one arrives at the general formula
\begin{equation}                                                        \label{eq:A03}
{\cal F}_k=(1/R)\int_{0}^1 P_k(x) e^{\sigma x^2}\,dx,
\end{equation}
where $\cal F$ is $S_k$ for even and is $Q_k$ for odd values of the index, and the function
$R(\sigma)$ is defined by Eq.~(\ref{eq:11}). In particular
\begin{equation}                                                        \label{eq:A04}
Q_1 =(1/R)\int_{0}^1 x e^{\sigma x^2}\,dx=\textstyle{\frac{1}{2}}(e^{\sigma}-1)/\sigma R.
\end{equation}
Using asymptotic expansion~(\ref{eq:12}) for $R$, one gets
\begin{widetext}
\begin{equation}                                                        \label{eq:A05}
Q_1=1/G=1-\frac{1}{2\sigma}-\frac{1}{2\sigma^2}-\frac{5}{4\sigma^3}-\frac{37}{8\sigma^4}
-\frac{353}{16\sigma^5}-\frac{4881}{32\sigma^6}-\frac{55205}{64\sigma^7}-
\frac{854197}{128\sigma^8}+\ldots
\end{equation}
Knowing $Q_1$, one can derive all the other moments $Q_k$ with the aid of the three-term
recurrence relation obtained from Eq.~(\ref{eq:08}) by setting there $b_k=Q_k$ and
$\lambda=0$. The same relation can be used to find the equilibrium order parameters $S_k$.
This is a head-to-tail procedure, where $S_0=1$ and $S_2$ is determined by the integral
\begin{equation}                                                        \label{eq:A06}
S_2=(1/2R)\int_{0}^1 \left(3x^2 -1\right) e^{\sigma x^2}dx.
\end{equation}
Taking the latter by parts one gets
\begin{displaymath}
S_2=\textstyle{\frac{3}{4}}\left[e^\sigma-R\right]/\sigma R.
\end{displaymath}
On comparison with Eq.~(\ref{eq:A04}), we find
\begin{displaymath}
S_2=\textstyle{\frac{3}{2}}\,Q_1-\textstyle{\frac{3}{4}}\,(3-2\sigma)/\sigma,
\end{displaymath}
that upon substituting asymptotic series~(\ref{eq:A05}), transforms into
\begin{equation}                                                        \label{eq:A07}
S_2=1-\frac{3}{2\sigma}- \frac{3}{4\sigma^2}-\frac{15}{8\sigma^3}
-\frac{111}{16\sigma^4}-\frac{1059}{32\sigma^5} -\frac{12243}{64\sigma^6}
-\frac{165615}{128\sigma^7}- \frac{2562591}{256\sigma^8}+\ldots
\end{equation}
\end{widetext}

\section{Evaluation of the correcting coefficients  $D_n$ in a general case \label{sec:AppB}}
Let us present the solution of Eq.~(\ref{eq:42}) in the form
\begin{equation}                                                        \label{eq:B01}
f_1^{(n)}=\psi_{0}g_{1}^{(n)}+u^{(n)},
\end{equation}
where the functions $g_{1}^{(n)}$ are rendered by formulas~(\ref{eq:45}) and are not corrected
with respect to the derivative $d\varphi_{1}/d x$. Substituting Eq.~(\ref{eq:B01}) in
(\ref{eq:42}) and taking into account Eqs.~(\ref{eq:45}), we get a recurrence sequence of
equations for the corrections $u^{(n)}$:
\begin{equation}                                                        \label{eq:B02}
\widehat{L}\,u^{(n)}=\widehat{V}\,u^{(n-1)}
+\widehat{\bm{J}}\psi_{0}\frac{(\bm{e}\cdot\bm{h})^n}{n!} \widehat{\bm{J}}\varphi_{1}\,.
\end{equation}
With allowance for the fact that function $\varphi_{1}^{(0)}$ depends only on $x$,
Eq.~(\ref{eq:B02}) rewrites as
\begin{displaymath}
\widehat{L}\,u^{(n)}=\widehat{V}\,u^{(n-1)}
+\frac{d}{dx}\,\left[\,\psi_0(1-x^2)\,\frac{(\bm{e}\cdot\bm{h})^n}{n!}
\frac{d\varphi_1}{dx}\right].
\end{displaymath}
Finally, making use of the relation
\begin{equation}                                                        \label{eq:B03}
\frac{d\varphi_1}{dx}=\frac{\lambda_1}{2\psi_0(1-x^2)},
\end{equation}
that follows from Eq.~(\ref{eq:18}), we get
\begin{equation}                                                        \label{eq:B04}
\widehat{L}\,u^{(n)}=\widehat{V}\,u^{(n-1)} +\frac{\lambda_1}{2}
\frac{d}{dx}\,\left[\,\frac{(\bm{e}\cdot\bm{h})^n}{n!}\right].
\end{equation}
In particular, at $n=1$ Eq.~(\ref{eq:B04}) takes the form
\begin{equation}                                                        \label{eq:B05}
\widehat{L}\,u^{(1)}=\frac{\lambda_1}{2}\frac{d}{dx}\,(\bm{e}\cdot\bm{h})
\end{equation}
Equations~(\ref{eq:B04}) are solved sequentially beginning from Eq.~(\ref{eq:B05}) by
expanding in a power series with respect to $x$. The right-hand sides of Eqs.~(\ref{eq:B04})
and (\ref{eq:B05}) are proportional to an exponentially small parameter $\lambda_1$. Just due
to that we did not take into account the corrections of the order $u^{(n)}$ when deriving
Eqs.~(\ref{eq:45}). However, the quantities
\begin{displaymath}
D_n = \left(\varphi_1\bigl|u^{(n)}\right) \qquad n=2,4..,
\end{displaymath}
have finite values. To show that, let us multiply Eq.~(\ref{eq:B04}) by $\varphi_1$ and
integrate. This yields
\begin{equation}                                                        \label{eq:B06}
\left(\varphi_1\bigl|\widehat{L}\,u^{(n)}\right)=
\left(\varphi_1\bigl|\widehat{V}\,u^{(n-1)}\right)+\frac{\lambda_1}{2}
\left(\varphi_1\bigl|\frac{d}{dx}\,\left[\,\frac{(\bm{e}\cdot\bm{h})^n}{n!}\right]\right).
\end{equation}
In the left part we make use of the fact that $\varphi_1$ is the left eigenfunction of the
operator $\widehat{L}$, in the right part the integrals are taken by parts and yield
\begin{eqnarray}                                                        \label{eq:B07}
\lambda_1 D_n=2\int_0^1(1-x^2)
u^{(n-1)}\left[\frac{d\varphi_1}{dx}\right]\frac{d}{dx}\,(\bm{e}\cdot\bm{h})\,dx \\%
\qquad -\lambda_1\int_0^1\frac{d\varphi_1}{dx}\frac{(\bm{e}\cdot\bm{h})^n}{n!}\,dx. \nonumber
\end{eqnarray}
Replacing the derivative $d \varphi_{1}/d x$ in the first term of the right-hand side with the
aid of Eq.~(\ref{eq:B03}), we arrive at the representation of the coefficient $D_n$ as
\begin{equation}                                                        \label{eq:B08}
D_n=\int_0^1 \frac{u^{(n-1)}}{\psi_0}\frac{d}{dx}\,(\bm{e}\cdot\bm{h}) dx- \int_0^1
\frac{d\varphi_1}{dx} \frac{(\bm{e}\cdot\bm{h})^n}{n!} dx.
\end{equation}
Since $\psi_0\propto\exp(\sigma x^2)$, the first integral in Eq.~(\ref{eq:B08}) can be
presented as an asymptotic series if the power expansion of the function $u^{(n-1)}$ in the
vicinity of $x=0$ is known. A closed form for the second integral can be found with the aid of
the table given in Eq.~(\ref{eq:51}), see section \ref{sec:3.1}.

As an example, we calculate the coefficient $D_2$. As from the addition theorem
\begin{displaymath}
(\bm{e}\cdot\bm{h})=\cos\theta\cos\beta+\sin\theta\sin\beta\cos\varphi,
\end{displaymath}
we seek the solution of Eq.~(\ref{eq:B05}) the sum
\begin{equation}                                                        \label{eq:B09}
u^{(1)}=\cos\beta\sum_k C_k^{(0)}x^k + \sin\beta e^{i\varphi} (1-x^2)^{\frac{1}{2}} \sum_k
C_k^{(1)}x^k.
\end{equation}
Here the upper index of the $C$ coefficients corresponds to the azimuthal number $m$ of the
spherical harmonic $e^{im\varphi}$. Operator $\widehat{L}$ now includes the azimuthal
coordinate and takes the form
\begin{eqnarray}
-\widehat{L}=(1-x^2)\frac{d}{d x^2}- \left[2\sigma x(1-x^2)+2x\right]\frac{d}{d x} \nonumber\\
+\left[2\sigma (3x^2-1)-\frac{m^2}{1-x^2}\right].                                 \nonumber
\end{eqnarray}
Substitution of expansion~(\ref{eq:B09}) in Eq.~(\ref{eq:B05}) leads to the set of equations
\begin{widetext}
\begin{equation}                                                        \label{eq:B10}
2\sigma(k+m+1)C_{k-2}^{(m)}- \left[k(k+1+2m+2\sigma)+m(m+1)+2\sigma\right]C_{k}^{(m)}+
(k+1)(k+2)C_{k+2}^{(m)}= N_{k}^{(m)},
\end{equation}
\end{widetext}
where $m=0,1$ and the numbers in the right-hand side are
\begin{displaymath}
N_{k}^{(0)}=\begin{cases} -1 & \text {for}\>\, k=0, \\ 0 & \text {for}\>\, k\neq0,
\end{cases} \qquad
N_{k}^{(1)}=\begin{cases} 1, & \text {for $k$ odd},\\ 0, & \text {for $k$ even}.\end{cases}
\end{displaymath}
In reality, one retains in expansion~(\ref{eq:B09}) only a finite number of terms so that
Eqs.~(\ref{eq:B10}) could be easily solved analytically by any computer algebra solver. In
terms of expansion~(\ref{eq:B09}) expression~(\ref{eq:B08}) at $n=2$ writes
\begin{eqnarray}                                                        \label{eq:B11}
D_2= \cos^2\beta \sum\limits_{k=0} C_{2k}^{(0)}\frac{(2k-1)!!}{2^k\sigma^k G} \\%
-\frac{1}{2}\sin^2\beta \sum\limits_{k=1} C_{2k-1}^{(1)}\frac{(2k-1)!!}{2^k\sigma^k G} \nonumber\\
-\frac{1}{6}- \frac{2G-3}{6G}P_2(\cos\beta). \nonumber
\end{eqnarray}
Since the coefficients $C$ found from Eq.~(\ref{eq:B10}) are functions of $\sigma$, one has to
perform in Eq.~(\ref{eq:B11}) asymptotic expansion. This gives finally
\begin{eqnarray}                                                        \label{eq:B12}
&& D_2=\frac{1}{4\sigma}+\frac{1}{4\sigma^2}+\frac{5}{8\sigma^3}+\frac{37}{16\sigma^4}+\ldots\\
&&-\sin^2\beta\left(\frac{1}{4}+\frac{1}{8\sigma}+\frac{1}{16\sigma^2}+ \frac{7}{64\sigma^3}+
\frac{19}{64\sigma^4}+ \ldots\right). \nonumber
\end{eqnarray}
As it should be, at $\beta=0$ this formula reduces to Eq.~(\ref{eq:52}) that was obtained for
a one-dimensional case. We remark, however, that in a tilted situation ($\beta\ne0$) the
coefficient $D_2$ acquires a contribution independent on $\sigma$ that assumes the leading
role. This effect is clearly due to admixing of transverse modes to the set of eigenfunctions
of the system, and it is just it that causes so a significant discrepancy between the
``zero-derivative'' approximation and the correct asymptotic expansion for $\chi^{(3)}$ curves
in Fig.~\ref{fig:05}. Evaluation of the coefficient $D_4$ is done according to the same scheme
and requires taking into account a number of the perturbation terms that makes it rather
cumbersome.

\section{Evaluation of integrals \label{sec:AppC}}
Before proceeding to the integrals (scalar products) in Eqs.~(\ref{eq:62})--(\ref{eq:66}) and
(\ref{eq:68}), let us consider the ``primitive'' ones
\begin{displaymath}
X_n=\left((\bm{e}\cdot\bm{h})^n\bigl|\psi_0\right), \qquad
Y_n=\left((\bm{e}\cdot\bm{h})^n\bigl|\psi_1\right).
\end{displaymath}
The functions $\psi_0$ and $\psi_1$ are originally defined in terms of the angle
$\theta=\arccos(\bm{e}\cdot\bm{n})$. Thus, before performing integration one needs to
transform both integrands to the same set of angles. Doing this with the aid of the addition
theorem for Legendre polynomials, one finds
\begin{widetext}
\begin{eqnarray}                                                        \label{eq:C01}
&& X_2=\textstyle{\frac{1}{3}}\,[\,2S_2P_2(\cos\beta)+1], \qquad
X_4=\textstyle{\frac{1}{35}}\,[\,8S_4P_4(\cos\beta)+20 S_2P_2(\cos\beta)+7],
                                                                        \nonumber \\[4pt]
&& X_6=\textstyle{\frac{1}{231}}\,[\,16S_6P_6(\cos\beta)+72
S_4P_4(\cos\beta)+110S_2P_2(\cos\beta )+33],
\end{eqnarray}
and
\begin{eqnarray}                                                        \label{eq:C02}
&& Y_1=Q_1\cos\beta\,, \qquad
Y_3=\textstyle{\frac{1}{5}}\,[\,2Q_3P_3(\cos\beta)+3
Q_1\cos\beta],
                                                                        \nonumber \\[4pt]
&& Y_5=\textstyle{\frac{1}{63}}\,[\,8 Q_5P_5(\cos\beta) +
28Q_3P_3(\cos\beta)+27Q_1\cos\beta],
\end{eqnarray}
where $\cos\beta=(\bm{n}\cdot\bm{h})$ and the parameters $S_k$ and $Q_k$ are the expansion
coefficients introduced by formulas~(\ref{eq:32}).

Now using the expressions for functions $f_0^{(n)}$ and $f_1^{(n)}$ derived in
Sec.~\ref{sec:3.1} one sees that the relevant integrals of Eqs.~(\ref{eq:63})--(\ref{eq:68})
are expressed in terms of $X_k$ and $Y_k$ as
\begin{equation}                                                        \label{eq:C03}
\left((\bm{e}\cdot\bm{h})\bigl|f_0^{(1)}\right)=X_2, \>\>
\left((\bm{e}\cdot\bm{h})\bigl|f_0^{(3)}\right)=\textstyle{\frac{1}{6}}X_4
-\textstyle{\frac{1}{2}}X_2^2, \>\>
\left((\bm{e}\cdot\bm{h})\bigl|f_0^{(5)}\right)=\textstyle{\frac{1}{120}}X_6
-\textstyle{\frac{1}{8}}X_4 X_2+\textstyle{\frac{1}{4}}X_2^3;
\end{equation}
\begin{equation}                                                        \label{eq:C04}
\left(\varphi_1\bigl|f_0^{(1)}\right)=Y_1\,, \>\>
\left(\varphi_1\bigl|f_0^{(3)}\right)=\textstyle{\frac{1}{6}}Y_3 -\textstyle{\frac{1}{2}}X_2
Y_1\,, \>\> \left(\varphi_1\bigl|f_0^{(5)}\right)=\textstyle{\frac{1}{120}}Y_5
-\textstyle{\frac{1}{12}}Y_3 X_2+\textstyle{\frac{1}{4}}X_2^2 Y_1 -\textstyle{\frac{1}{24}}X_4
Y_1\,;
\end{equation}
\begin{eqnarray}                                                        \label{eq:C05}
&& \left(\varphi_1\bigl|f_1^{(2)}\right)=\textstyle{\frac{1}{2}}X_2-Y_1^2+D_2\,, \quad
\left((\bm{e}\cdot\bm{h})\bigl|f_1^{(2)}\right)=\textstyle{\frac{1}{2}}Y_3 -Y_1X_2+D_2Y_1\,,
                                                                        \nonumber\\[4pt]
&& \left(\varphi_1\bigl|f_1^{(4)}\right)=\textstyle{\frac{1}{24}}X_4
-\textstyle{\frac{1}{3}}Y_3Y_1+\textstyle{\frac{1}{2}}X_2Y_1^2 +D_4
+D_2\left(\varphi_1\bigl|f_1^{(2)}\right)\,,                            \nonumber \\[4pt]
&& \left((\bm{e}\cdot\bm{h})\bigl|f_1^{(4)}\right)=\textstyle{\frac{1}{24}}Y_5
-\textstyle{\frac{1}{6}}Y_3X_2 -\textstyle{\frac{1}{6}}X_4 Y_1 +\textstyle{\frac{1}{2}}X_2^2
Y_1+ D_4Y_1+D_2\left((\bm{e}\cdot\bm{h})\bigl|f_1^{(2)}\right).
\end{eqnarray}
\end{widetext}

\end{document}